\documentstyle[psfig]{mn}

\footnotesize
\newdimen\digitwidth    
\setbox0=\hbox{\rm0}
\digitwidth=\wd0
\catcode`!=\active
\def!{\kern\digitwidth}
\normalsize

\newcommand{\axp}{AXP\,J1810$-$197\,}

\title[Polarized radio emission from a magnetar]{Polarized radio emission
from a magnetar}
\author[M.~Kramer et al.]
{
M.~Kramer$^1$\thanks{Email: mkramer@jb.man.ac.uk},
B.~W.~Stappers$^2$, A.~Jessner$^3$, A.~G.~Lyne$^1$, C.A.~Jordan$^1$
\\
$^1$University of Manchester,
Jodrell Bank Observatory, Macclesfield, Cheshire, SK11~9DL, UK\\
$^2$Stichting ASTRON, Postbus 2, 7990 AA Dwingeloo, The Netherlands\\
$^3$MPI f{\"u}r Radioastronomie, Auf dem H{\"u}gel 69, 53121 Bonn, Germany
}

\date{accepted for publication in MNRAS}
\begin{document}

\maketitle
\newcommand{\setthebls}{
}

\setthebls

\begin{abstract} 
  We present polarization observations of the radio emitting magnetar
  \axp. Using simultaneous multi-frequency observations performed at
  1.4, 4.9 and 8.4 GHz, we obtained polarization information for
  single pulses and the average pulse profile at several epochs. We
  find that in several respects this magnetar source shows
  similarities to the emission properties of normal radio pulsars
  while simultaneously showing striking differences.  The emission is
  nearly 80-95\% polarized, often with a low but significant degree of
  circular polarization at all frequencies which can be much greater
  in selected single pulses. The position angle swing has a low
  average slope of only 1 deg/deg, deviating significantly from an
  S-like swing as often seen in radio pulsars which is usually
  interpreted in terms of a rotating vector model and a dipolar
  magnetic field. The observed position angle is consistent at all
  frequencies while showing significant secular variations. On average
  the interpulse is less linearly polarized but shows a higher degree
  of circular polarization. Some epochs reveal the existence of
  non-orthogonal emission modes in the main pulse and systematic 
  wiggles in the PA swing, while the
  interpulse shows a large variety of position angle values. We
  interprete many of the emission properties as propagation effects in
  a non-dipolar magnetic field configuration where emission from
  different multipole components is observed.
\end{abstract}


\section{Introduction}

Until very recently, radio emission from neutron stars was only detected from
radio pulsars. While pulsars were discovered almost 40 years ago, and their
radio emission has been studied very extensively (Lyne \& Smith
2006)\nocite{ls06}, the processes responsible for the observed radiation
properties are not understood. It is remarkable that essentially identical
emission features are observed over a wide range of parameter space,
i.e.~covering nearly four orders of magnitude in spin period, about seven
orders of magnitude in spin-down rate and about six orders of magnitude in
surface magnetic field strength. Recent discoveries now allow us to study
other, more exotic neutron stars in the radio regime and to contrast their
emission properties with those of normal pulsars. This promises to shed light
on the conditions that are required in a neutron star magnetosphere to produce
coherent and highly polarized emission.

On one hand, a new class of radio transient sources named {\em
  Rotating Radio Transients} (RRATs) has been identified as rotating
neutron stars (McLaughlin et al.~2006)\nocite{mll+06}.  These emit
radio emission in an irregular, bursty fashion for a total timespan of
less than one second per day. Despite some of these sources possibly
being radio pulsars with unusual pulse energy distributions
(Weltevrede et al. 2006)\nocite{wsrw06}, it is possible that some of
these sources provide an observational link between radio pulsars and
the so-called {\em magnetars}. Magnetars are considered to be slowly
rotating neutron stars visible as bursting X-ray and gamma-ray
sources, characterized by long spin-periods of 5-12 seconds and a
rapid spin-down (see the detailed review by Woods \& Thompson
2006)\nocite{wt06}.  The observed X-ray and gamma-ray luminosities
are usually orders of magnitude larger than would be expected if they
were powered by the spin-down energy.  Interpreting the spin-down as
being caused by magnetic dipole radiation, extremely large magnetic
field strengths, typically $>10^{14}$~G, are inferred.  According to
the model by Duncan \& Thompson (1992)\nocite{dt92a}, it is the
spontaneous decay of the very large magnetic fields of these sources
which powers the high energy emission.  Originally, this model was
proposed to explain the soft gamma-ray repeaters (SGRs), but
observations of the so called Anomalous X-ray pulsars (AXPs) indicate
that both types of objects share many properties and are therefore
considered as a unified class of magnetars.
  
Studying the possible relationship between radio pulsars and magnetars is
intriguing, since highly magnetized radio pulsars have been discovered (e.g.
McLaughlin et al.~2003)\nocite{msk+03} that exhibit spin properties which are
essentially identical to those of some magnetars. Searches for X-ray emission
for some of these highly magnetized pulsars has indeed been successful (Kaspi
\& McLaughlin 2005)\nocite{km05}.  On the other hand, magnetars were thought
to be radio quiet. Reports of pulsed radio emission from magnetars at low
radio frequencies at or near 100 MHz (e.g.~Shitov et al.~2000\nocite{spk00})
could not be confirmed with other telescopes (e.g.~Lorimer \& Xilouris
2000\nocite{lx00}).  This situation has changed quite dramatically with the
discovery of a unique anomalous X-ray pulsar.

The AXP XTE J1810$-$197 was discovered by chance during observations of the
known magnetar SGR 1806-20 by Ibrahim et al. (2004)\nocite{ims+04}.  A clearly
periodic signal with a period of 5.54 seconds was detected with the Rossi
X-ray Timing Explorer (XTE). A check of archival XTE data showed that the
source had been present in galactic bulge scans since February 2003 and had
likely gone into outburst sometime late in 2002. Using these archival
observations it was possible to obtain a timing solution and determine a
period derivative of $1.15\times10^{-11}$~s~s$^{-1}$ which appeared to be
unsteady and varying by a factor of several within a few months of the
outbursts. Using standard radio pulsar methods, such spin-down implies a
magnetic field strength of about $2.6\times10^{14}$ Gauss.  The
luminosity calculated from the spin-down, $\dot{E} \sim
4\times10^{33}$~erg~s$^{-1}$, is about two orders of
magnitude less than the observed X-ray burst luminosity. Since the period and
magnetic field strength are very typical of the known AXPs, the source was
indeed identified to be a member of this class.  However, unusually for the
AXPs and SGRs, it shows large variations in the X-ray flux and its sudden
appearance has meant that it has been referred to as the first transient AXP.

What is also unusual for AXPs is that it was detected as a radio source by
Halpern at al.~(2005).\nocite{hgb+05} Observations taken as part of the VLA
MAGPIS survey on 2004 January 1 at 1.4 GHz revealed a bright 4.5 mJy source at
the precise location of the AXP. From other archival observations it was not
possible to determine whether the flux density of the source was decaying,
indicating it may be associated with the outburst, or whether it might be a
nebula powered by the AXP.

Camilo et al. (2006a)\nocite{crh+06} reported observations of the
radio emission using the Parkes radio telescope at 1.4 GHz on 2006
March 17 which, remarkably, revealed a strong detection of pulsations
with the known 5.54~s period.  This represents the first clear
evidence of radio pulsations from a magnetar.  The radio pulsations
are found to be remarkable in many ways, the pulse profile being variable and formed
from sharp and narrow pulse components.  The profile shows significant
evolution as a function of frequency with new components appearing at
other pulse longitudes. Individual pulses are seen from almost every
rotation and they show very narrow bright sub-pulses which are highly
linearly polarised.

Flux density variations of almost a factor of two are also seen between
successive observations and are inconsistent with the behaviour expected from
interstellar scintillation.  Furthermore radio pulsations from this source
were also detected at other frequencies near 1.4 GHz and these revealed that
the pulsar seemed to have an unusually flat spectrum of $S\propto\nu^{-0.5}$
or flatter. Subsequent observations at higher radio frequencies have shown that
it is the brightest known neutron star at frequencies above 20 GHz. Very
recently, Camilo et al.~(2006b) \nocite{ccr+06} presented additional evidence
for the time-variability of the source, and also showed that the radio
emission appears to be becoming weaker.  Interestingly, simultaneous X-ray and
radio observations suggest that the radio and X-ray profiles are aligned.

In this paper we report on the first results of simultaneous multi-frequency
full-polarisation observations conducted at radio frequencies
around 1.4 GHz, 4.9 GHz and 8.4 GHz. We present examples of single pulse
polarization but concentrate here more on the properties of the
average pulse profiles, while an analysis of the single pulse properties 
will be reported elsewhere.

\section{Observations}

Observations were made using the 76-m Lovell radio-telescope at Jodrell
Bank Observatory of the University of Manchester, UK, the 94-m
equivalent Westerbork Synthesis Radiotelescope (WSRT) in the
Netherlands, and the 100-m radio-telescope of the Max-Planck Institute
for Radioastronomy (MPIfR) at Effelsberg, Germany.  Details of the
observing set-ups can be found in Gould \& Lyne (1998)\nocite{gl98},
Karastergiou et al.~(2001)\nocite{khk+01} and van Leeuwen 
et al.~(2003)\nocite{vsrr03} while details of the observing
sessions are summarized in Table~\ref{tab:obs}.

\begin{table}
\caption{\label{tab:obs} Summary of observing sessions.}
\begin{tabular}{cclcc}
\hline
\hline
Session & Epoch (MJD ) & Telescope & Freq. (MHz) & BW (MHz) \\
\noalign{\smallskip}
\hline
\noalign{\smallskip}
1 & 53886 & Lovell & 1418 & 32  \\
  &       & WSRT   & 4901 & 160  \\
  &       & Effelsberg & 8350 & 1000 \\
\noalign{\smallskip}
2 & 53926 & Lovell & 1418 &  32 \\
  &       & Effelsberg & 8350 & 1000 \\
\noalign{\smallskip}
3 &  53933 & Lovell & 1418 & 32 \\
  &        & WSRT   & 4896 & 160 \\
  &        & Effelsberg & 8350 & 1000 \\
\noalign{\smallskip}
4 &  53937 & WSRT   & 4896 & 160 \\
\noalign{\smallskip}
5 & 53938 & Lovell & 1418 & 32 \\
  &       & WSRT   & 4896 & 160 \\
  &       & Effelsberg & 8350 & 1000 \\
\noalign{\smallskip}
6 &  53939 & Lovell & 1418 & 32 \\
  &        & WSRT   & 4896 & 160 \\
\noalign{\smallskip}
7 & 53942 & Lovell & 1418 & 32 \\
  &       & Effelsberg & 4850 & 500 \\
  &       & Effelsberg & 8350 & 1000 \\
\noalign{\smallskip}
8 & 53944 & Lovell & 1418 & 32 \\
  &       & Effelsberg & 4850 & 500 \\
  &       & WSRT       & 4896 & 160 \\
  &       & Effelsberg & 8350 & 1000 \\
\noalign{\smallskip}
\hline
\end{tabular}
\end{table}

\subsection{Observing systems \& Calibration procedures}

At Jodrell Bank, we employed the 1.4-GHz receiving system as described in
Karastergiou et al.~(2001). In short, after mixing to an intermediate frequency
(IF), the two orthogonal circularly polarized signals were fed into a
$32\times 1.0$ MHz-channel filterbank system, with incoherent de-dispersion
performed in hardware producing all four Stokes parameters.  A dispersion
measure of 178 cm$^{-3}$ pc (Camilo et al.~2006a) was used for all observations.
Due to hardware
constraints in the de-disperser, the number of bins per period is limited. In
order to overcome this restriction and to allow for sufficient time
resolution, we sampled the single pulses in three sets. Small gaps in the
total period were needed for hardware read-out.  At the beginning of each
observation, the pulse phase of the first sampled bin was chosen such that the
gaps were located in off-pulse regions. In total, we cover the period
with 1536 phase bins, corresponding to an effective resolution of 3.6ms.

At Westerbork, we mostly used the 4.9-GHz receiving system with a
bandwidth of 80 MHz distributed evenly in 10 MHz sections across a total
band of 160 MHz.  The two linear polarisations from all 14 telescopes
were added by taking account of the relative geometrical and
instrumental phase delays between them. Phase differences between the
two linear polarisations were corrected by making observations of a
known polarised calibrator. The baseband data were then passed to the
PuMa pulsar backend which formed a digital filterbank of the four
Stokes parameters \cite{vkv02}. De-dispersion and folding were
carried out offline. While the raw data have a 
sampling time of 0.8 ms,
the data presented here were re-sampled to an effective resolution of 6.5 ms.

At Effelsberg, we mostly used the 8.35-GHz HEMT receiver installed in the
secondary focus but also used occasionally the 4.85-GHz receiver. The latter
system was described in detail by Karastergiou et al.~(2001), providing a
total bandwidth of 500 MHz. The 8.35-GHz system is very similar but provides a
total bandwidth of 1 GHz. As at 4.85 GHz, the wide-bandwidth circularly
polarized IF signals are fed into a multiplying polarimeter, detected in the
focus cabin and are digitized by fast voltage-to-frequency converters. All
Stokes parameters are recorded with 1024 phase bins of 5.3 ms duration, leaving a
short time for data read out in each period. During the first 50 phase bins, a
calibration signal was switched on, injecting the signal of a noise diode into
the front-end waveguide. The known polarization characteristics of this 100\%
linearly polarized signal allowed for reliable calibration of each single
pulse. In the profiles shown later, this signal is blanked out to show the
astronomical signal only. As the
observing frequencies used were sufficiently high to neglect dispersion
smearing, the signals could be sampled without applying de-dispersion. With a
dispersion smearing of 6.5 ms and 2.5 ms at 4.85 GHz and 8.35 GHz, respectively,
the effective resolutions are 8.4 ms and 5.9 ms.

All data were calibrated (see e.g.~Gould \& Lyne 1998\nocite{gl98},
von Hoensbroech \& Xilouris 1997,\nocite{hx97}, Mitra et
al.~2003\nocite{mwkj03}, Edwards \& Stappers 2004,\nocite{es04}) and
converted into the European Pulsar Network (EPN) data format (Lorimer
et al.~1998)\nocite{ljs+98}. Polarization quality was cross-checked by
observing standard well-known pulsars such as PSR B1929+10 and PSR
B1933+16. In one instance (JBO data, session 1), polarization data
were rejected because proper calibration was not possible. As we will
demonstrate later, simultaneous observations at different
frequencies, using different telescopes of different mounts, different
polarimeters and data acquisition systems produced results that agreed
extremely well, giving confidence in the quality and reliability of
the observed degree of linear and circular polarization and the
measured position angle.

In addition to the simultaneous multi-frequency polarization observations
reported here, we also performed frequent and regular timing observations of
\axp.  Our timing parameters are consistent with the most recent ones reported
by Camilo et al.~(2006b) within their uncertainties, so that we do not present
our solution here. Unfortunately, our polarization observations are not
sampled densely enough to allow us to find possible correlations between
observed variations in the spin-down and secular changes in polarization
properties that we report in the following. Since there are significant
variations in the spin-down, profiles measured at different epochs were
aligned using both the polarization and total power information. In the latter
case, it proved useful to inspect the observed intensity on a logarithmic
scale, allowing studies of emission at very low intensity levels in order to
line up the emission windows.

\section{Results}

The multi-dimensional character of our dataset allows us to investigate the
properties of the radio emission of \axp{} as a function of frequency and time
on short and long-time scales. We will therefore discuss the various results
of our observations, both here and in subsequent papers, by focusing on
different aspects of the radio emission.  We will contrast our results to the
wealth of observational data collected for the polarized emission of normal
radio pulsars.  While the polarization properties of radio pulsars are only
rather poorly understood (e.g.~Lorimer \& Kramer 2005)\nocite{lk05}, they are
well studied observationally. This applies also to the simultaneous
multi-frequency properties of radio pulsars. This has been achieved with the
same network of telescopes as used for this study, in a collaboration known as
the European Pulsar Network (EPN). The results of multi-frequency polarization
properties of pulsars have been presented in particular in a series of papers by
Karastergiou et al.~(2001, 2002, 2003)\nocite{khk+01,kkj+02, kjk03}
or Johnston et al.~(2006)\nocite{jkw06}.

\subsection{Individual pulses}

While we will discuss the properties of individual pulses in much more detail
in a forthcoming publication, some of the properties of the average pulse
profiles to be discussed below can only be understood by realizing some rather
unusual single pulse characteristics, at least when compared to normal radio
pulsars. As already pointed out by Camilo et al.~(2006a,b), single pulses are
often very bright and easily detectable at high radio frequencies. Such
observations are facilitated not only by a flat flux density spectrum, but
also by very narrow pulse widths. A pulse can consist of a single or a few
very bright spikes.

\begin{figure*}
\centerline{
\begin{tabular}{c@{\hspace{-4em}}c}
\psfig{file=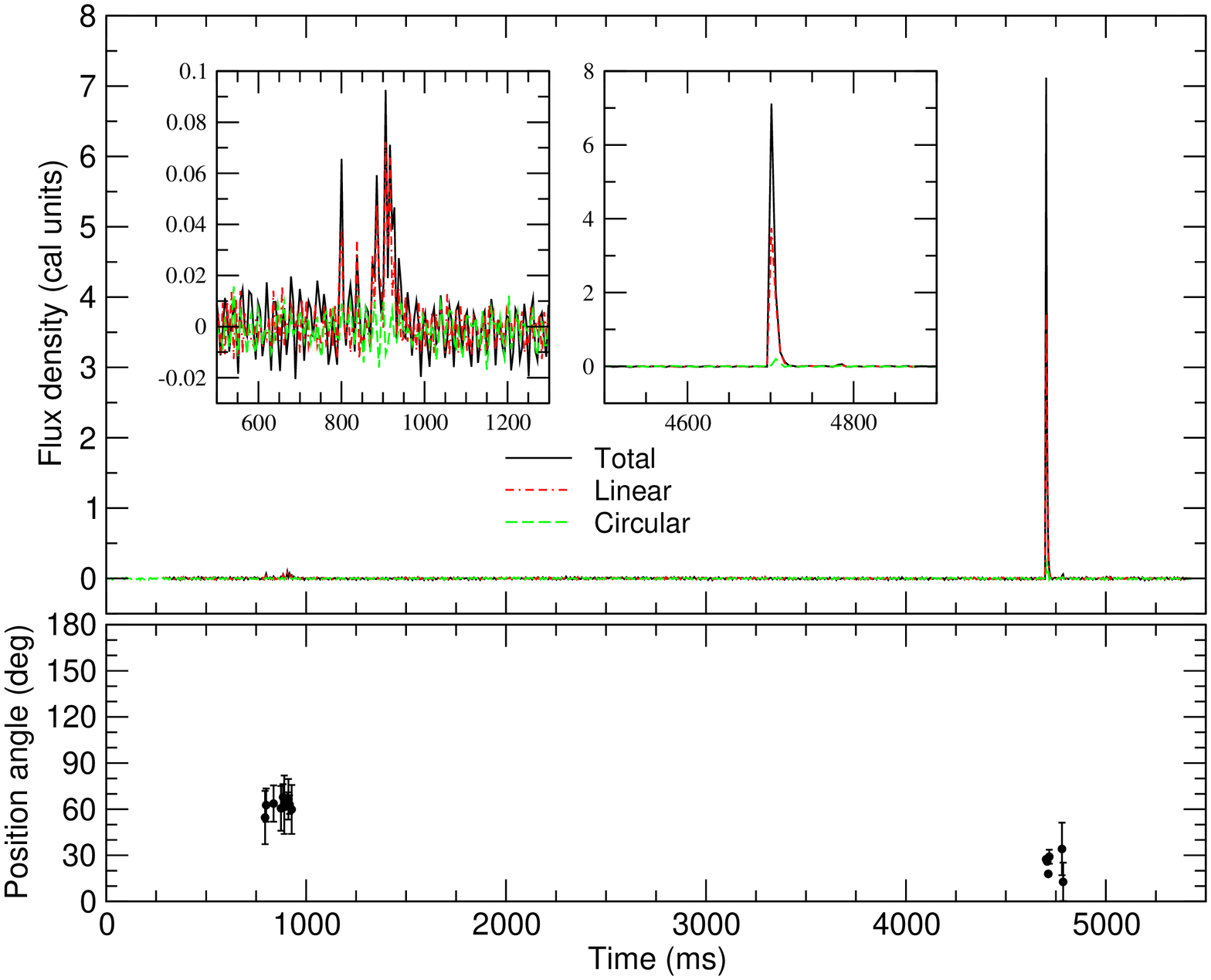,width=11cm,angle=-90} &
\psfig{file=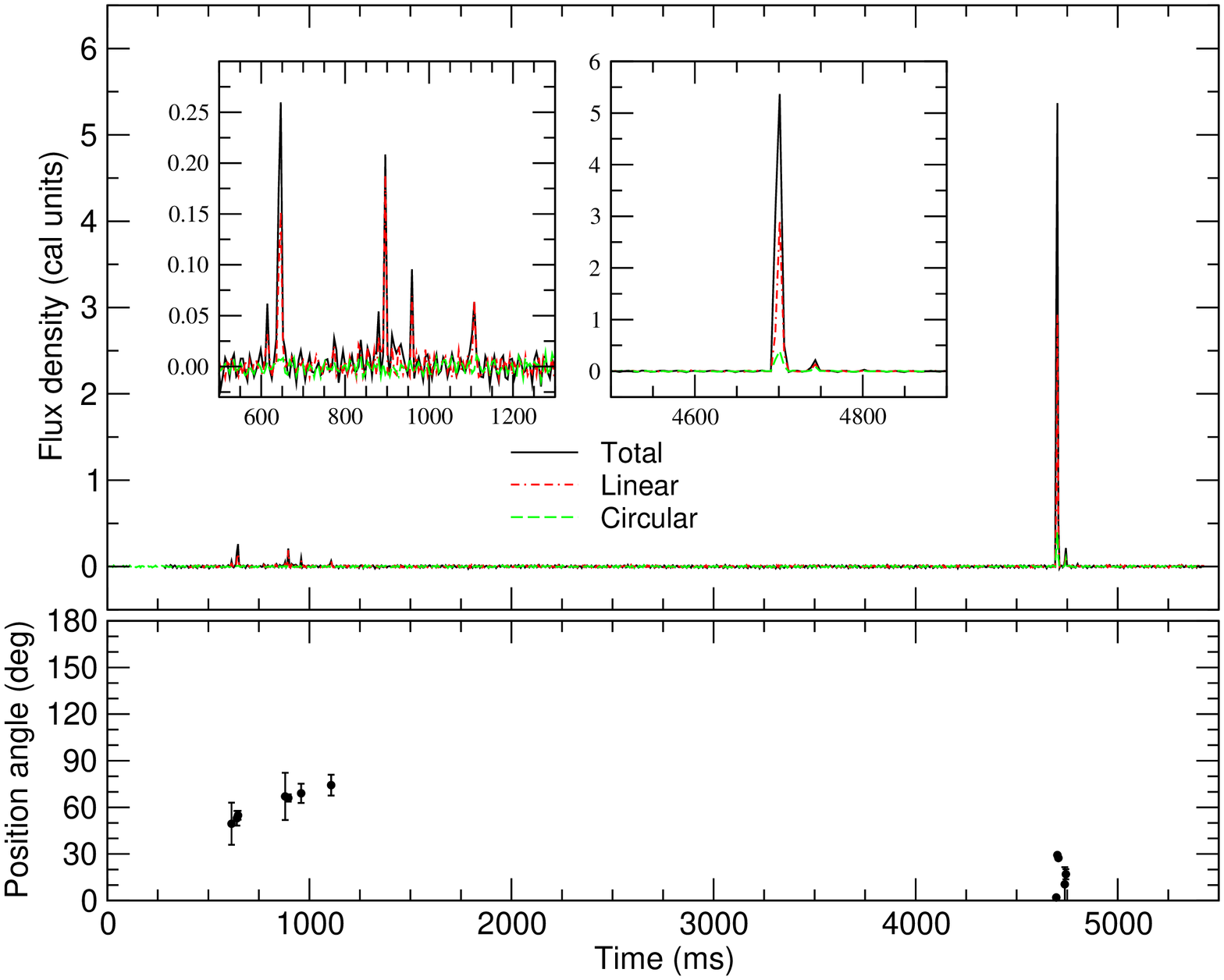,width=11cm,angle=-90} \\
\psfig{file=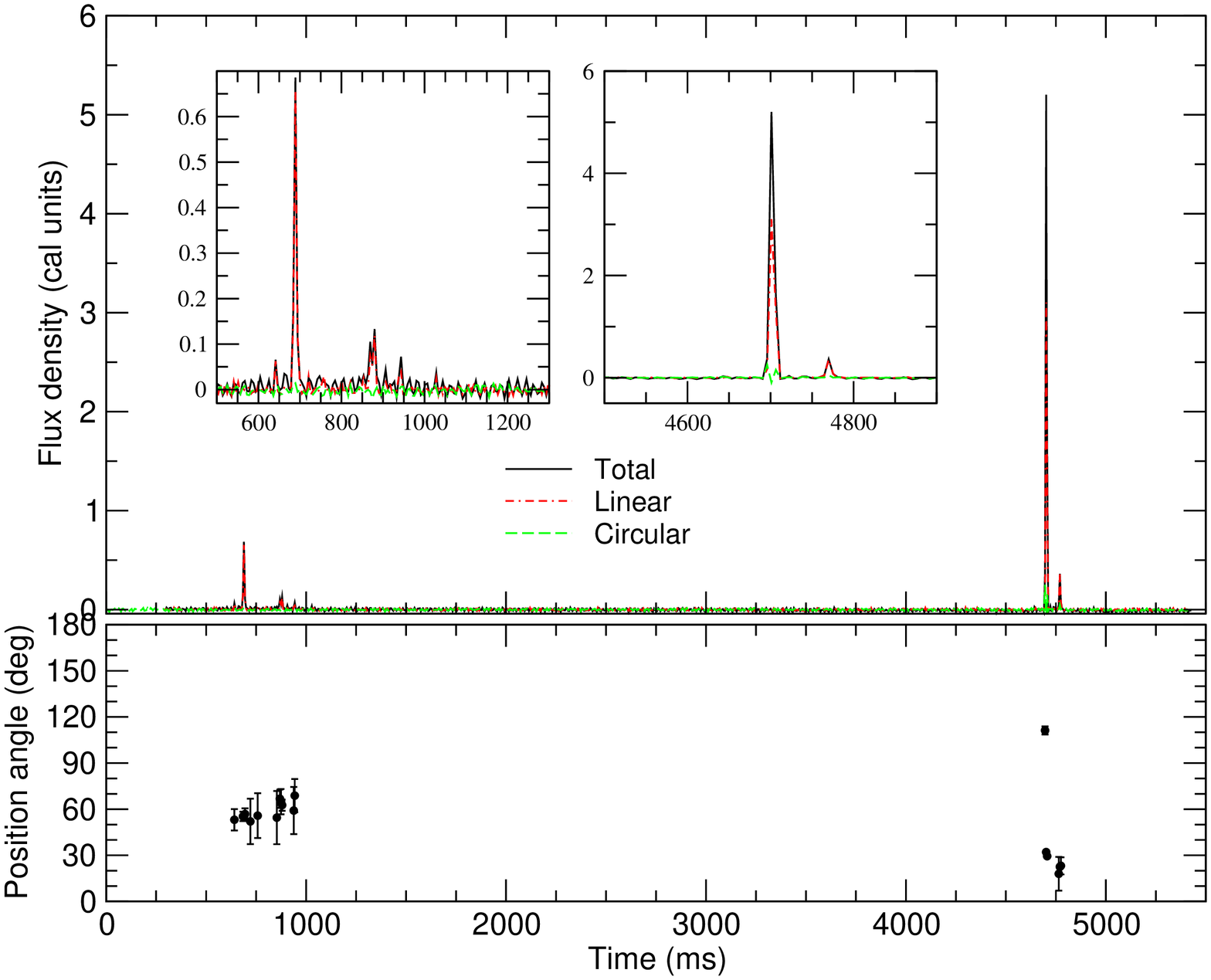,width=11cm,angle=-90} &
\psfig{file=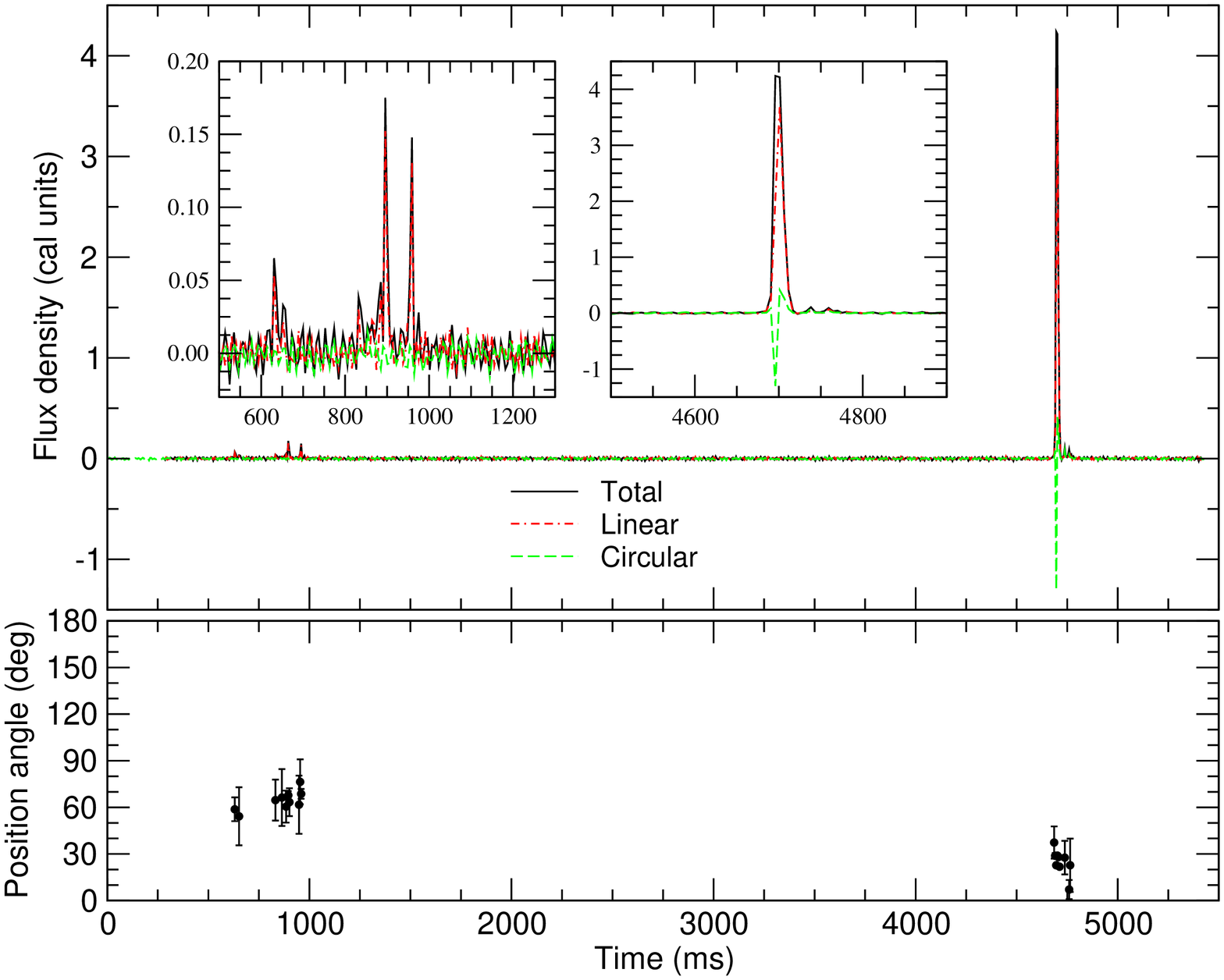,width=11cm,angle=-90}
\end{tabular}
}

\caption{\label{fig:biggies} The four strongest pulses
  as measured during session 3 at 8.4 GHz in full polarization. Insets shown
  enlarged regions of the MP (left) and IP (right).  In all cases, the
  components in the IP region (around time 4800~ms) shows stronger emission
  than those in the MP region (around time 1000~ms) as also found for the next
  six strongest pulses.  Note the large degree of circularly polarized
  emission seen with different handedness.}
\end{figure*}

We find all individual pulses to be very highly elliptically polarized,
typically between 80\% and 95\%. The linear polarization component is by far
the dominant one, but significant amounts of circular polarization are
detected as well. Typically, the degree of circular polarization
appears to be much larger in the emission component that we will refer
to as the {\em interpulse} (IP), discussing its possible
interpretation later in this paper.  We will refer to the collection
of the other prominent pulse features occurring about 4 seconds
earlier as the {\em main pulse} (MP). In
Figure~\ref{fig:biggies}, we select the four brightest single pulses
as seen at 8.4 GHz.

\begin{figure*}
\centerline{
\begin{tabular}{c@{\hspace{-4em}}c}
\psfig{file=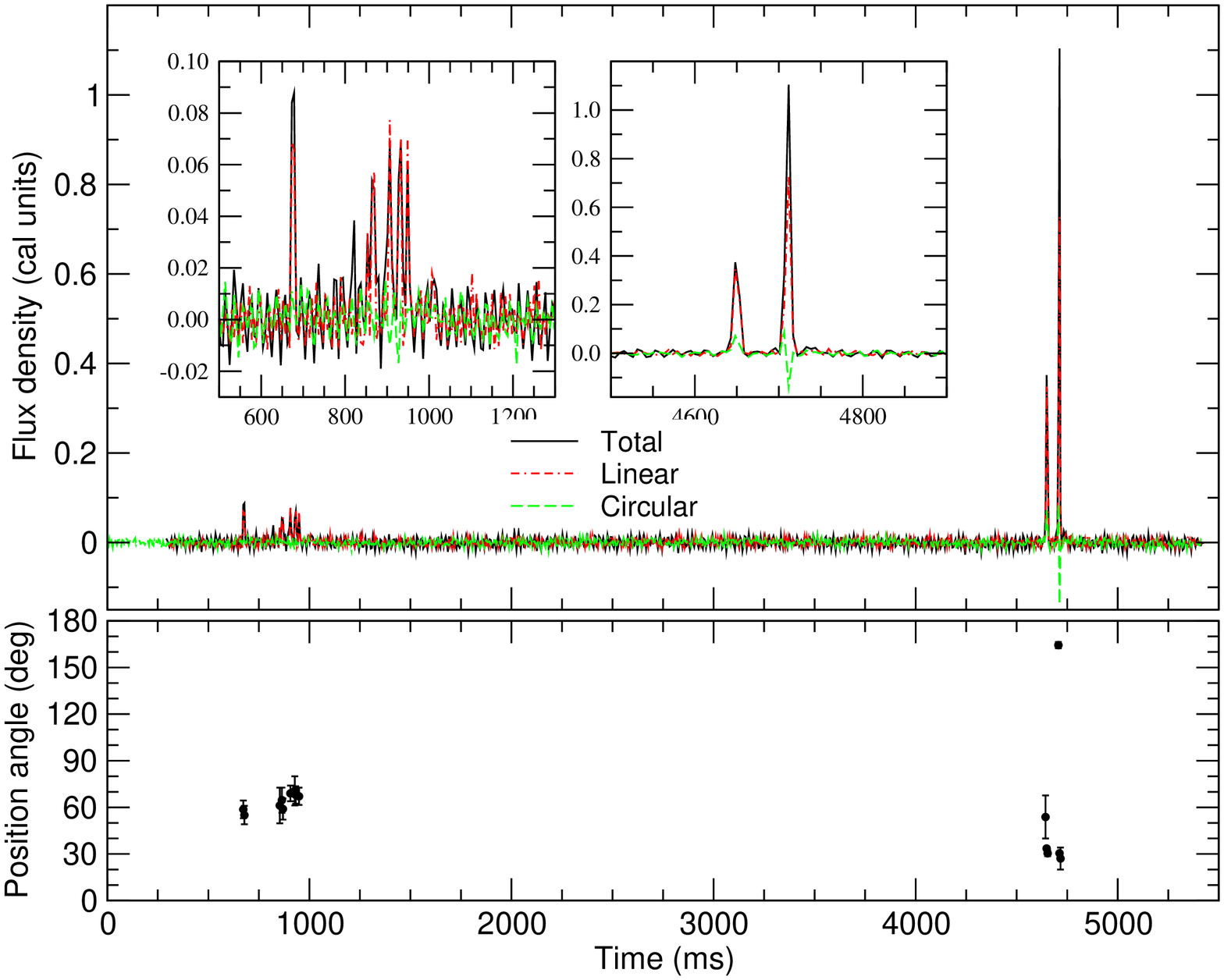,width=11cm,angle=-90} &
\psfig{file=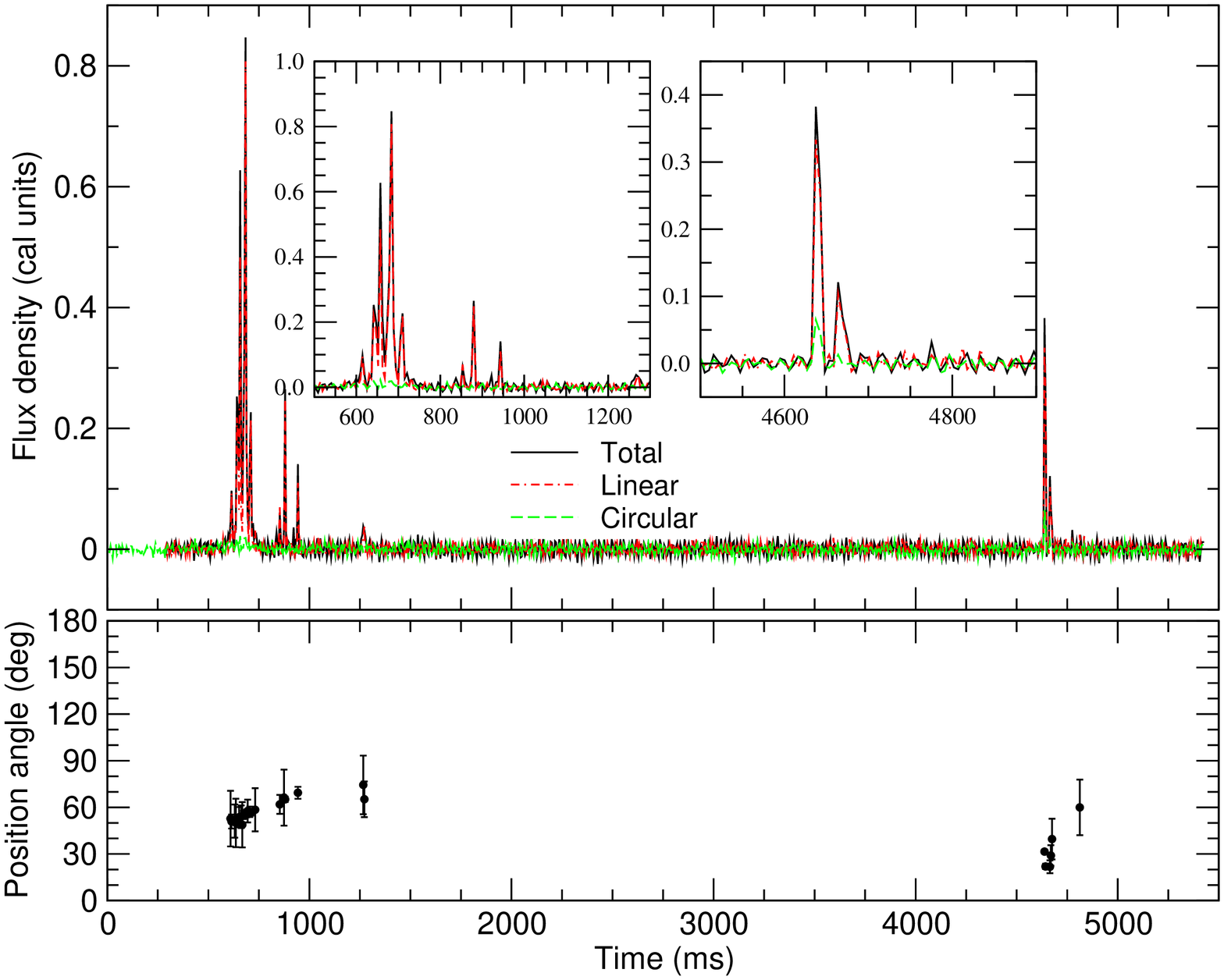,width=11cm,angle=-90} \\
\psfig{file=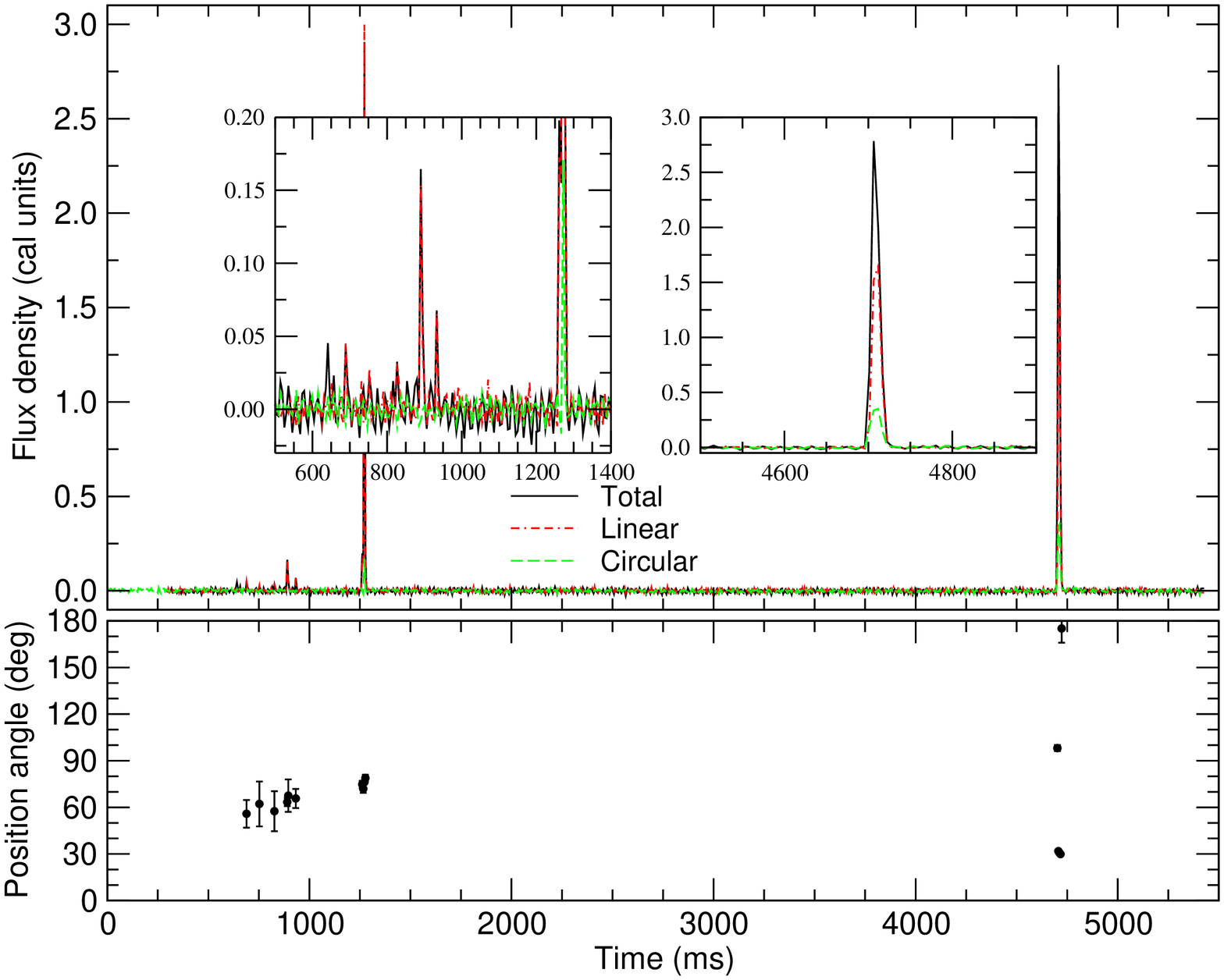,width=11cm,angle=-90} &
\psfig{file=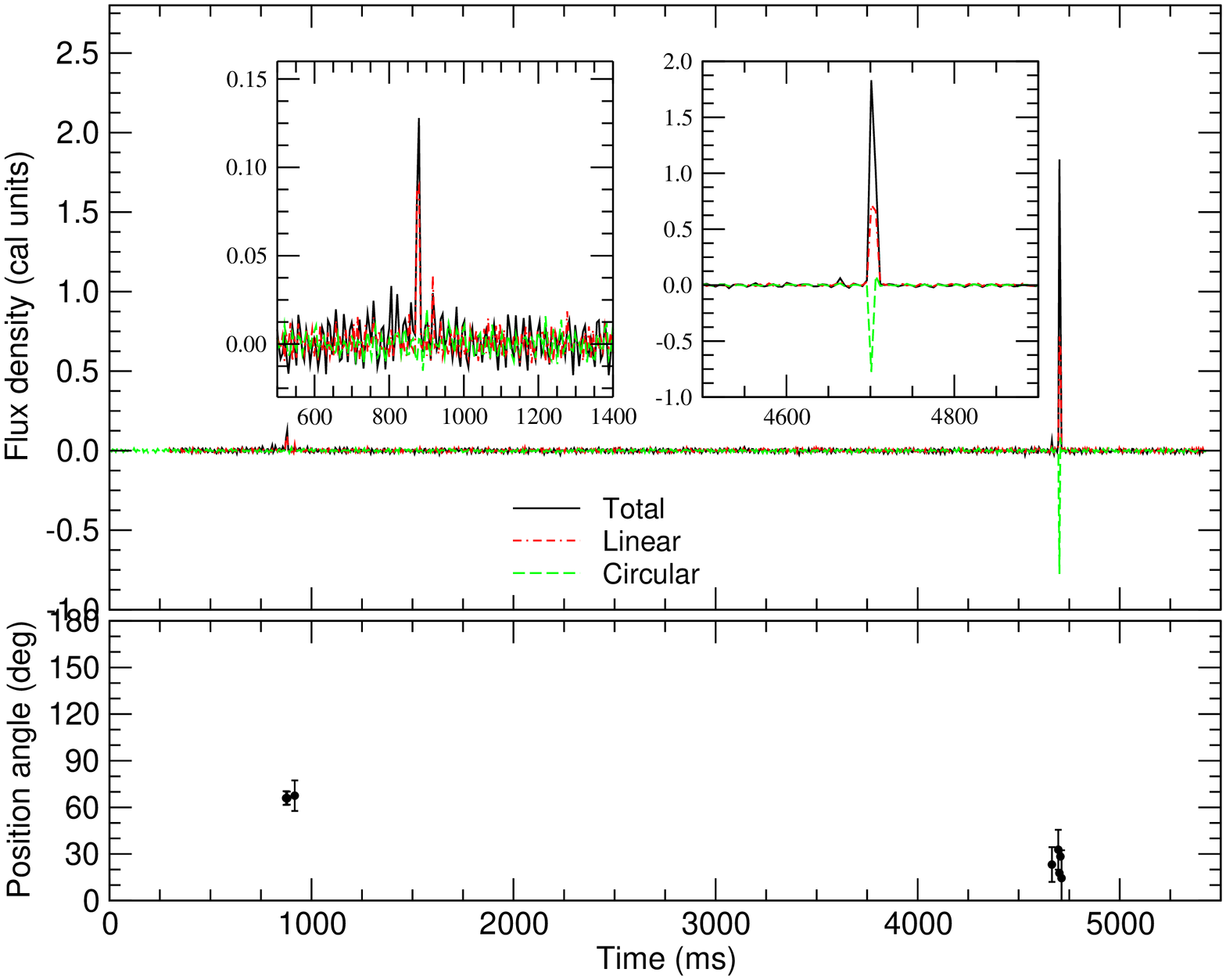,width=11cm,angle=-90}
\end{tabular}
}

\caption{\label{fig:variety} A sample of further pulses from the same
data set used for the previous figure, demonstrating the wider variety
of pulses appearing at pulse phases forming the MP. Note that the MP
is nearly 100\% linearly polarized while the IP shows some strong
circular polarization of changing handedness.}
\end{figure*}

In these particular observations, the brightest pulses were occurring
predominantly at the location of the IP. The MP is much broader and
consists of several, rather independent emission components. Hence,
the variety of single pulses seen for these pulse phases is much
greater. This is also demonstrated in Figure~\ref{fig:variety} where
we select four pulses to show the MP complexity and the rather strong
degree of circular polarization occurring under the IP.  Generally, the
IP is somewhat less strongly polarized but shows a higher degree of
circular polarization. The sense of the circular polarization
component can also change from pulse to pulse. For a single pulse, the
PA in the MP appears to be flattish although a small but significant
slope is noticeable if the PA can be measured over a sufficient range
of pulse longitudes. This is consistent with an average PA slope in
the MP of 1 deg/deg.

\subsection{Average pulse shapes at multiple frequencies}

The variation of the integrated
pulse shape on very short time scales as reported by
Camilo et al.~(2006b) is also visible in our data. While the total power
profile changes, the degree of linear polarization remains extremely high.  At
the same time, changes in circular polarization in both degree of polarization
and handedness are observed on short and long time scales.  The average
position angle swing is typically stable over the time scale of days, but we
do detect different (non-orthogonal) polarization modes which modify this
statement to some extent (see below). 

In Figure~\ref{fig:align}, we present the polarization of the pulse
profile observed at various frequencies during the observing
sessions. We also include multi-frequency data that are observed only
quasi-simultaneously (i.e.~within periods of a few hours) as we find
that the polarization features are much more persistent than the total
power profiles which change on time-scales of 10 minutes (cf.~Camilo
2006b). The properties of single pulses seen simultaneously at
different frequencies will be discussed in a forthcoming paper.

In the first session (Fig.~\ref{fig:align}a), the source was very
strong at all frequencies and single pulses were easily detected. No
polarisation is available in this session at 1.4 GHz. The profile
shows significant evolution with frequency, such that the trailing MP
components become more distinct towards 8.4 GHz.  Significant
left-handed circular polarization is detected under the strong peak
consistently at 4.9 GHz and 8.4 GHz, while the linear polarization
exceeds 90\%.  The position angle is identical at the different
frequencies, showing deviations from a smooth curve by exhibiting a
deep 'dip' at the longitude apparently separating the first and second
components.

In general the swing does not follow a S-like curve as would be expected in
a rotating vector model \cite{rc69a} but is rather shallow. The average slope
is rather modest, $d\Psi/d\Phi=1$ deg/deg covering only 45 deg in PA over the
whole MP emission window. The steepest slope is measured during session 1 in
the leading pulse component with a value of $d\Psi/d\Phi=3.5$ deg/deg.  As we
will discuss later in some detail, the PA swing of the IP changes between
observing sessions but typically shows a larger, negative slope.

In the second observing session (Fig.~\ref{fig:align}b), the frequency
evolution of the pulses occurring at the MP is much stronger with the greatest
complexity exhibited at the highest frequencies.  Remarkably, the IP was not
detected at 1.4 GHz but appeared clearly at 8.4 GHz. This trend is repeated
during the third session (Figure~\ref{fig:align}c), where the IP goes from
just barely detectable at 1.4 GHz to become the dominant component at 8.4 GHz,
indicating a much flatter spectrum of the IP relative to the MP. Where
detected, the IP shows consistently significant left-handed circular
polarization, but the degree of linear polarization is less than in the MP
which remains more than 90\% linearly polarized.  Circular polarization of the
same sense is weak but still detected at 1.4 GHz under the leading MP
component, while the later MP components are stronger at higher frequencies,
allowing the detection of circular polarization of the opposite sense.  The PA
swing has significantly changed from the first observations.  This change is
discussed in detail in the next section.

During session 5, the IP is still very weak at 1.4 GHz, but in sessions 7 and
8 (Figure~\ref{fig:align}e,f) the MP and IP are essentially of similar
strength at all frequencies. Still, this should not be taken as a sign that
the MP is becoming stronger with time.  Instead, it demonstrates the
significant time variability of the IP and its spectrum in particular. In any
case, while the general polarization characteristics of the IP have not
changed, the PA swing is now significantly different.  Interestingly, the
circular polarization that is clearly detected in the MP at 4.9 and 8.4 GHz
appears to be predominately left-handed, but both frequencies consistently
show a sense-reversal from right-hand to left-hand under the IP.

\begin{figure*}
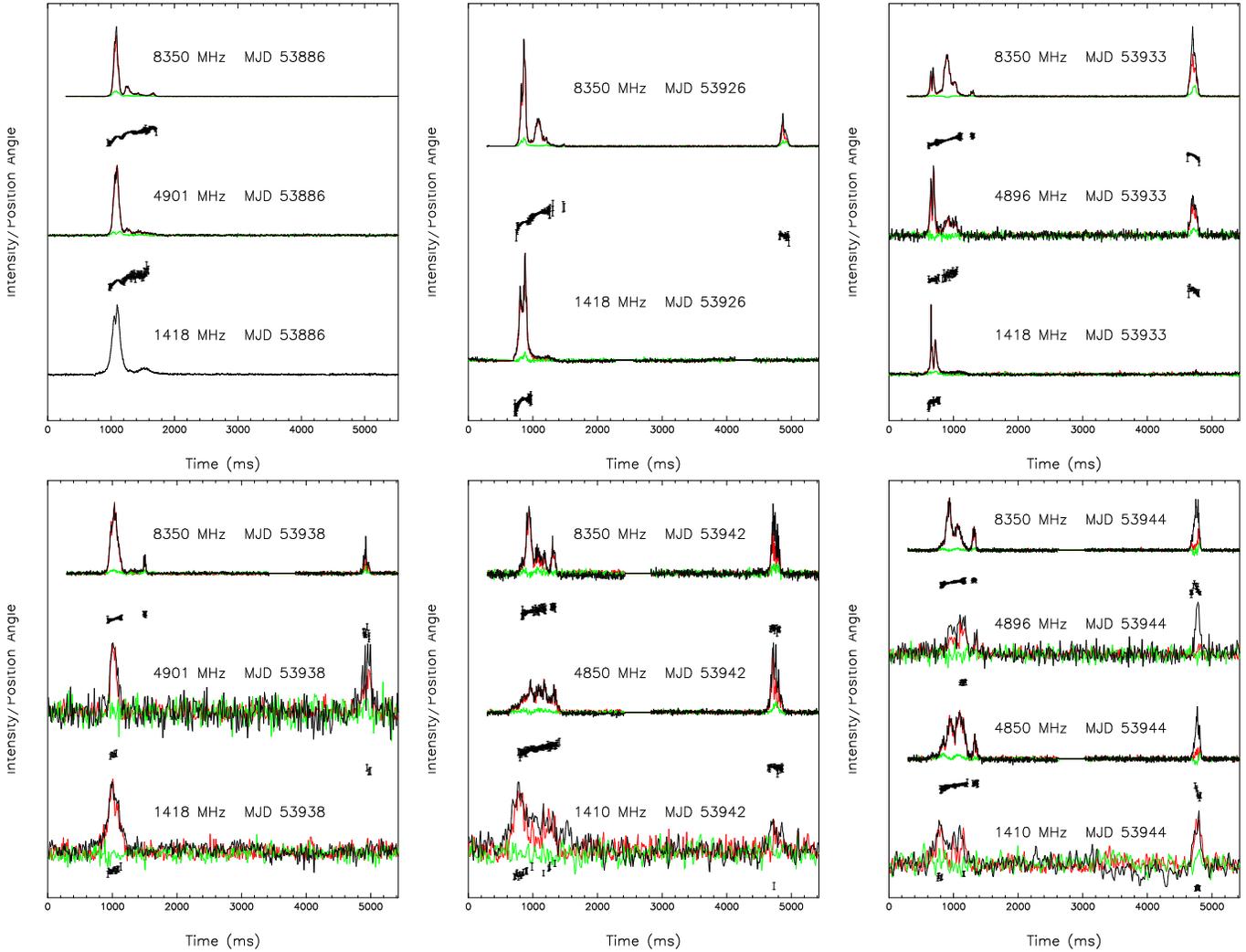


\begin{center}
\begin{tabular}{ccc}
\psfig{file=fig3a.ps,width=5.7cm} & 
\psfig{file=fig3b.ps,width=5.7cm} & 
\psfig{file=fig3c.ps,width=5.7cm} \\
\psfig{file=fig3d.ps,width=5.7cm} & 
\psfig{file=fig3e.ps,width=5.7cm} & 
\psfig{file=fig3f.ps,width=5.7cm} 
\end{tabular}
\end{center}

\caption{\label{fig:align}
  Average pulse profiles in full polarization observed at different
  frequencies for observing sessions 1,2,3,5,7 and 8. Total power (black) and
  linearly (red) and circularly (green) polarized power are shown together
  with the measured position angle for each epoch and frequency.
 The scale of the shown
  position angle swing is identical for all pulse profiles of a given epoch
  and presented here only to allow for a comparison of its general shape and
  extent. A quantitative comparison is made in Figures 7 and 8.  }
\end{figure*}

\subsection{Average pulse profiles at multiple epochs}

\begin{figure*}
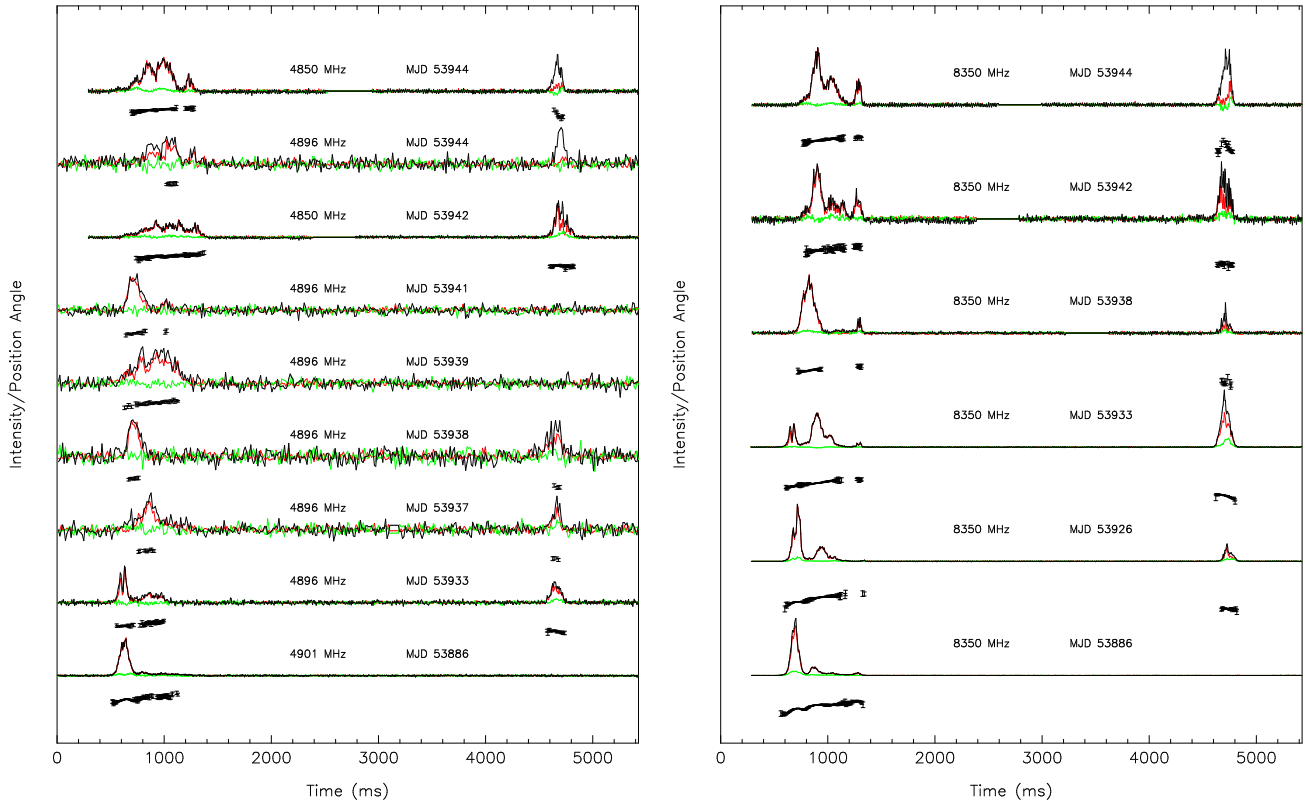


\begin{center}
\begin{tabular}{cc}
\psfig{file=fig4a.ps,width=8.4cm} & 
\psfig{file=fig4b.ps,width=8.4cm} 
\end{tabular}
\end{center}

\caption{\label{fig:epoch}
Comparison of the average pulse profiles in full polarization observed 
over several weeks at 4.9 GHz and 8.4 GHz. 
Total power (black) and
  linearly (red) and circularly (green) polarized power are shown together
  with the measured position angle for each epoch and frequency.
The scale of the shown
  position angle swing is again identical for all pulse profiles of a given epoch
  and presented here only to allow for a comparison of its general shape and
  extent. A quantitative comparison is made in Figures 7 and 8.  
}
\end{figure*}

 In order to study the siginificant pulse shape changes in more detail
 with respect to the polarization properties, we present these with
 additional pulse profiles measured at 4.9 GHz and 8.4 GHz in
 Figure~\ref{fig:epoch} over a period of several weeks.

The most striking feature in both plots is clearly the coming and
going of certain emission components with time. Moreover, while the MP
stays essentially completely linearly polarized, the circular
polarization remains low but mostly left-handed in the strong MP
components. In contrast, the circular polarization in the IP is
changing handedness, reflecting the variation seen in the individual
pulses. Moreover, the IP shows a typically much smaller degree of
polarization in general, with the PA swing clearly changing slope and
shape. It is also notable that the MP component that was clearly
dominant in the early observing session has disappeared towards the
later epochs. 

\begin{figure}

\begin{center}
\begin{tabular}{c}
\psfig{file=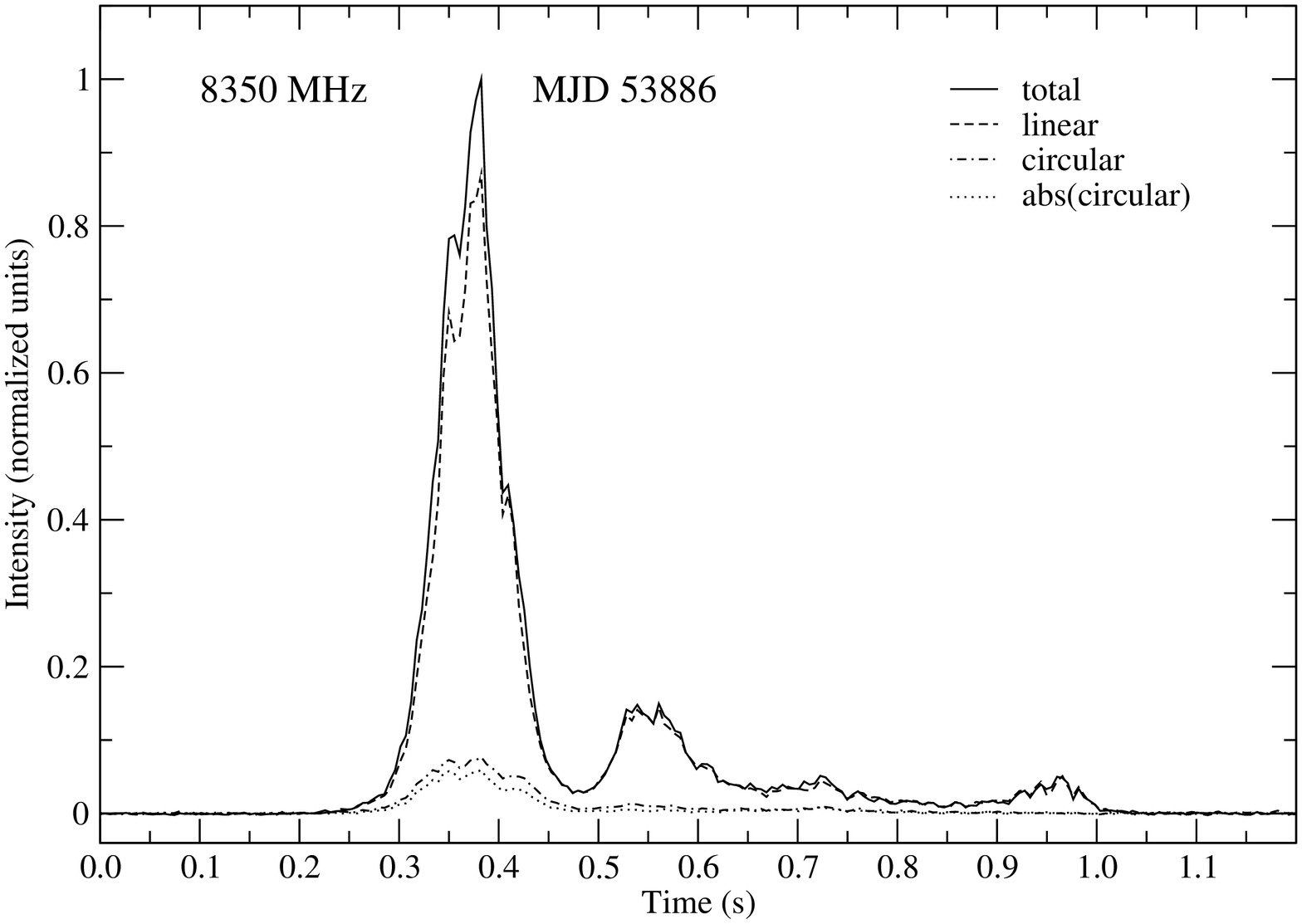,width=8cm,angle=-90} \\
\psfig{file=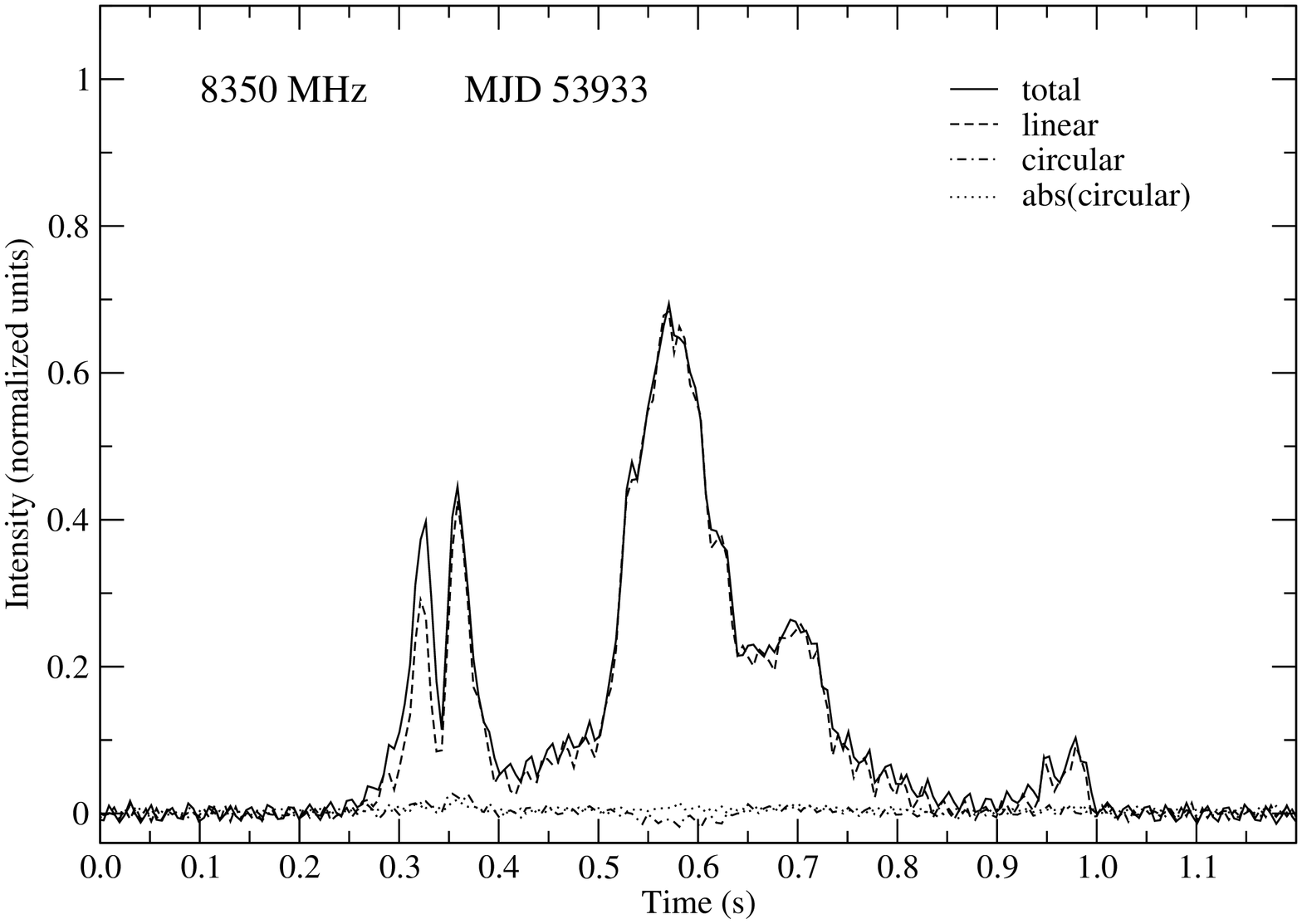,width=8cm,angle=-90} 
\end{tabular}
\end{center}

\caption{\label{fig:absV1}
Polarization of main pulse as measured at 8.4 GHz during sessions 1 and 3.
The leading component which disappears toward the later epochs is less
linearly polarized than the trailing components. It also shows a significant
amount of circular polarization which becomes weaker at later epochs.
The degrees of polarization for these two epochs are 
(total: $89.4\pm0.2$\%, linear: $88.8\pm0.2$\%, circular: $10.9\pm0.1$\%,
absolute circular: $10.4\pm0.1$\%) and 
(total: $92.8\pm0.7$\%, linear: $92.6\pm0.7$\%, circular: $0.8\pm0.4$\%,
absolute circular: $1.7\pm0.2$\%), respectively.
Corresponding values observed at the
lower frequencies are consistent.
}
\end{figure}

Inspecting the polarization properties in detail, one notices that the first
component of the main pulse is less polarized than the later components (see
Figure~\ref{fig:absV1} and caption). As we will see further on, this is a
region where a larger variety of position angles are observed, but it also
conincides with a region of significant amount of average circular
polarization.  It is then interesting to note that the degree of circular
polarization is almost identical to the amount obtained when adding the
absolute value of circular intensity to form the average profile. That
indicates that the variety of circular polarization seen in the
spiky single pulses for a wide range of pulse phases essentially cancels
out and only a lower level of persistent circular polarization remains. At later
epochs, this first component seems to gradually disappear, and the only highly
linearly polarized components remain.

\begin{figure}

\begin{center}
\begin{tabular}{c}
\psfig{file=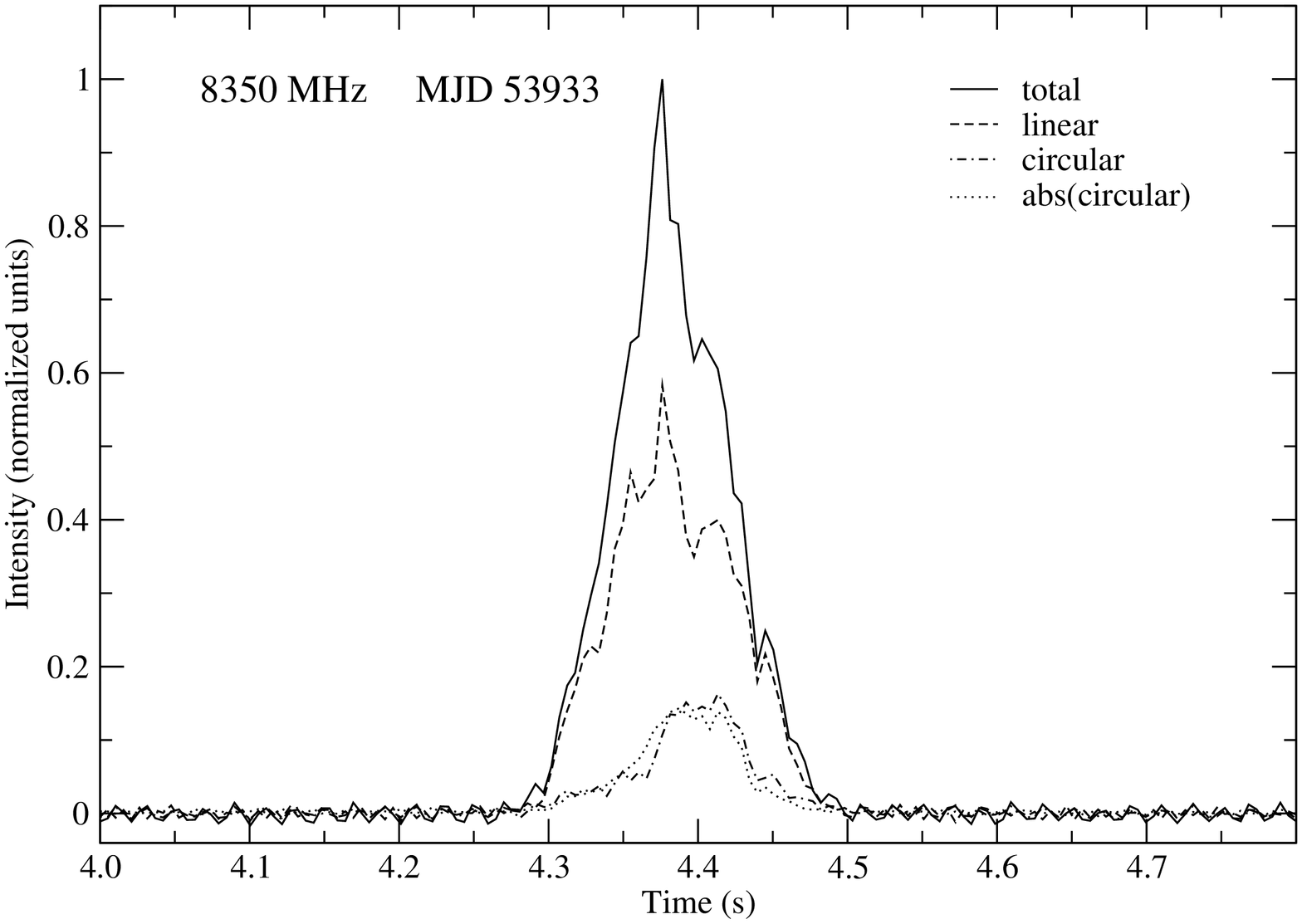,width=8cm,angle=-90} \\
\psfig{file=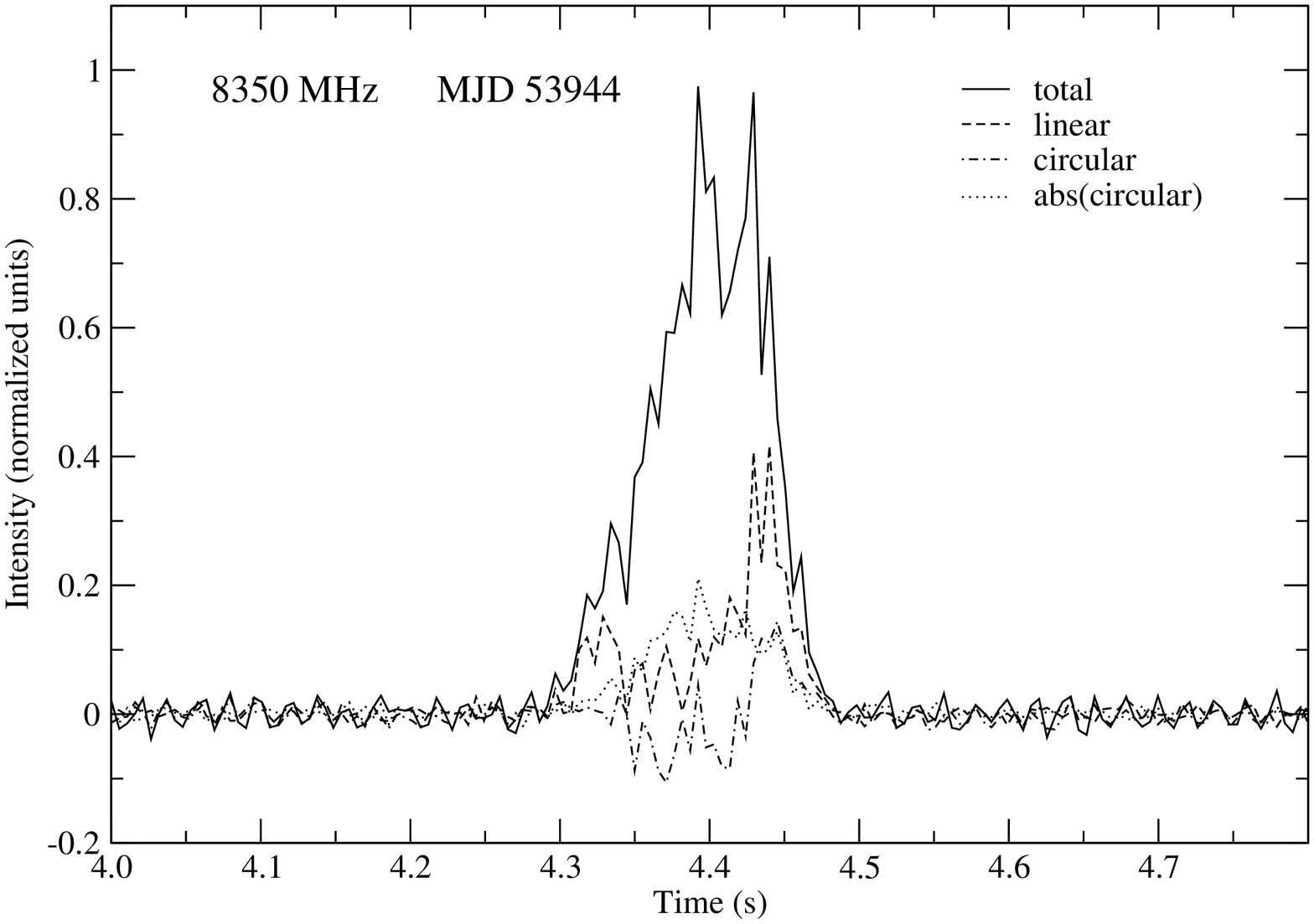,width=8cm,angle=-90} 
\end{tabular}
\end{center}

\caption{\label{fig:absV2}
Polarization of interpulse as measured at 8.4 GHz during sessions 3 and 8.
Significant amounts of positive circular is seen during the first epoch,
while the circular intensity is more structured and with sense reversal
during the later epoch.
The degrees of polarization for these two epochs are 
(total: $69.8\pm0.5$\%, linear: $68.1\pm0.5$\%, circular: $16.0\pm0.3$\%,
absolute circular: $16.0\pm0.3$\%),
(total: $27.6\pm1.0$\%, linear: $26.2\pm1.1$\%, circular: $0.0\pm1.0$\%,
absolute circular: $8.9\pm0.7$\%), respectively.
Corresponding values observed at the
lower frequencies are consistent.
}
\end{figure}

The situation is somewhat different for the interpulse, which is significantly
less polarized than the main pulse (see Figure~\ref{fig:absV2} and caption). A
variety of circularly polarized single pulses leaves a significant but
changing degree of circular polarization.  This conincides with a lower degree
of linear polarization with a large variety of
position angle values.

\subsection{Position angle swings}

\begin{figure*}

\begin{center}
\begin{tabular}{cc}
\psfig{file=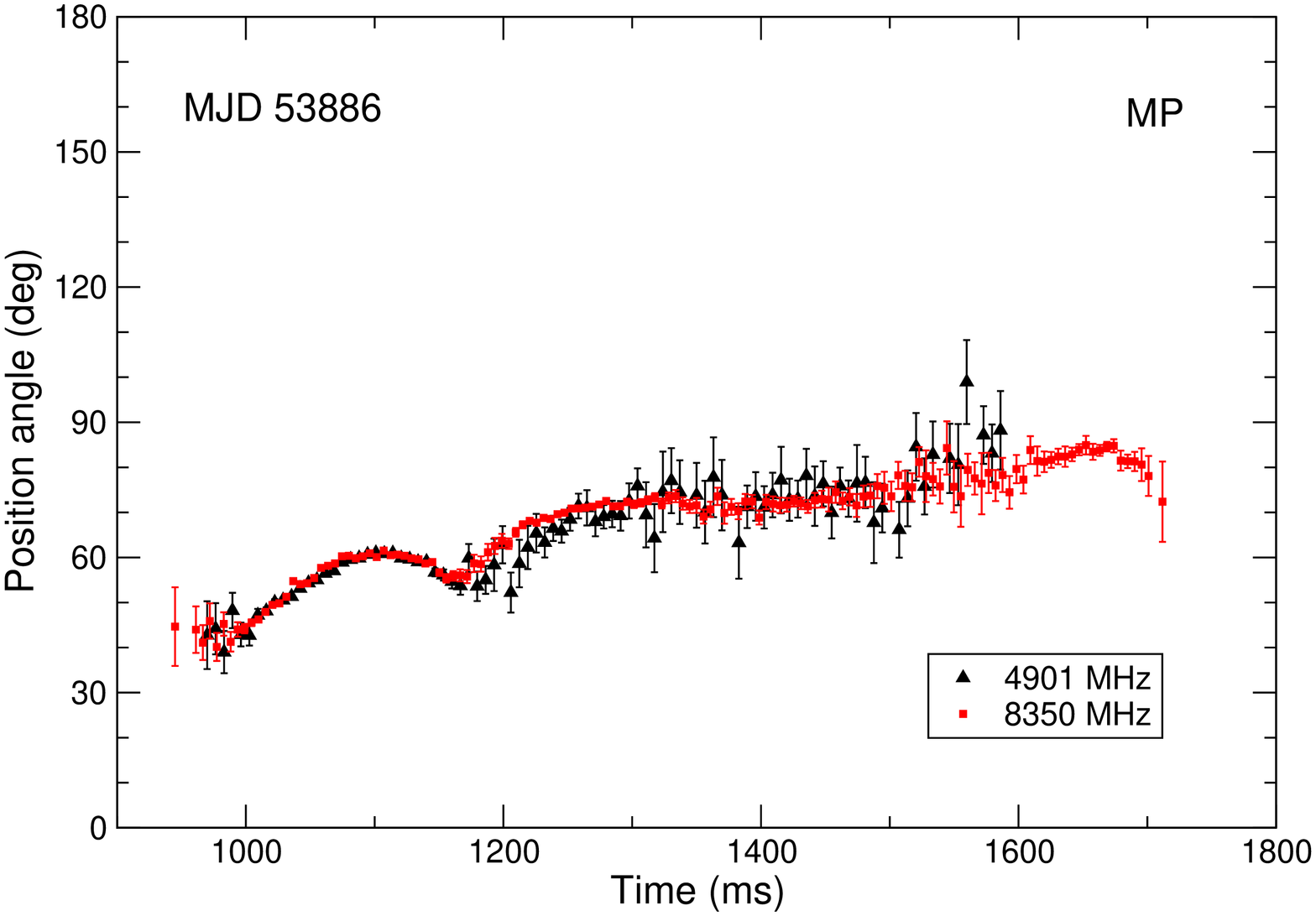,width=8cm,angle=-90} & 
\psfig{file=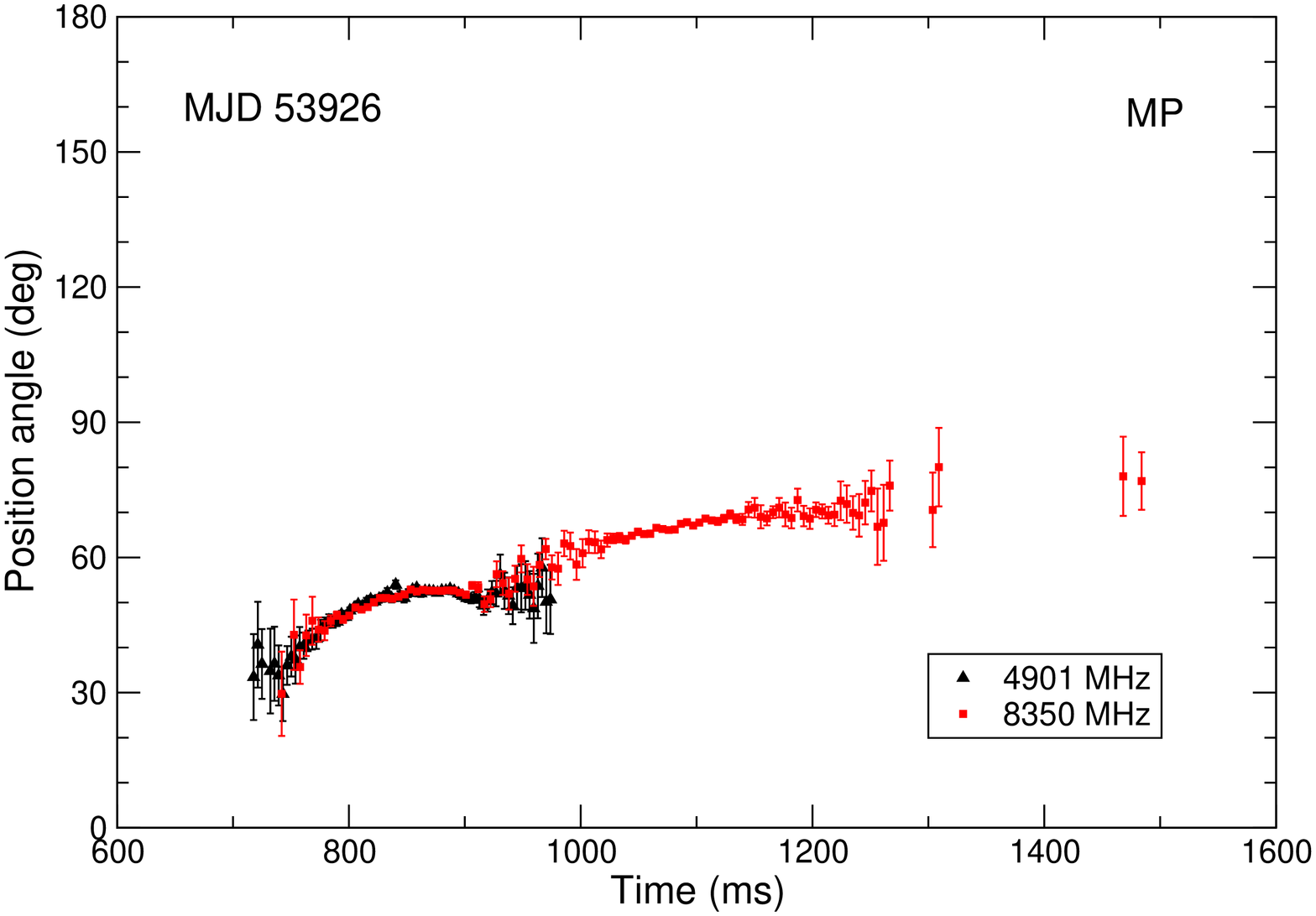,width=8cm,angle=-90} \\
\psfig{file=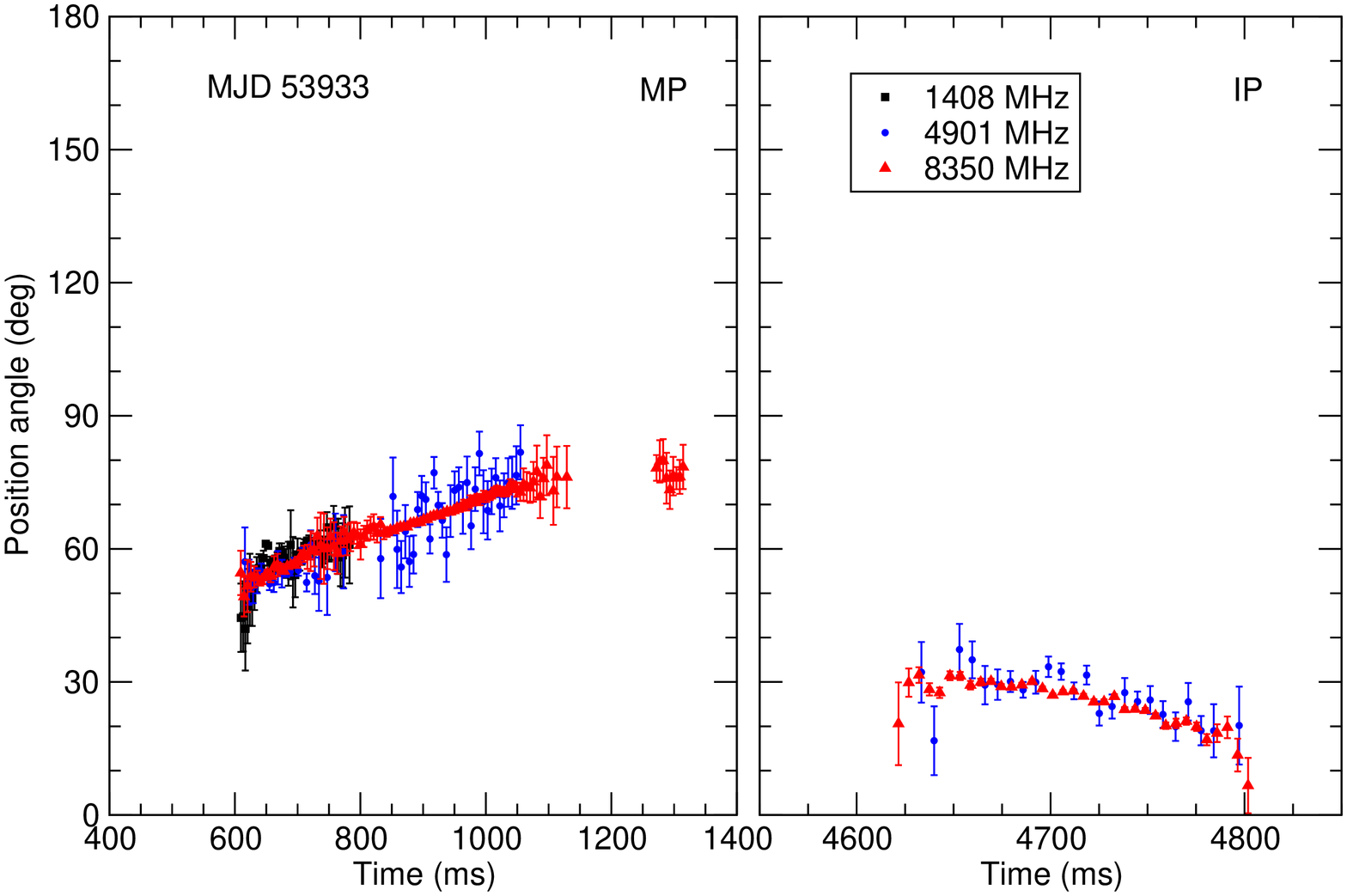,width=8cm,angle=-90} &
\psfig{file=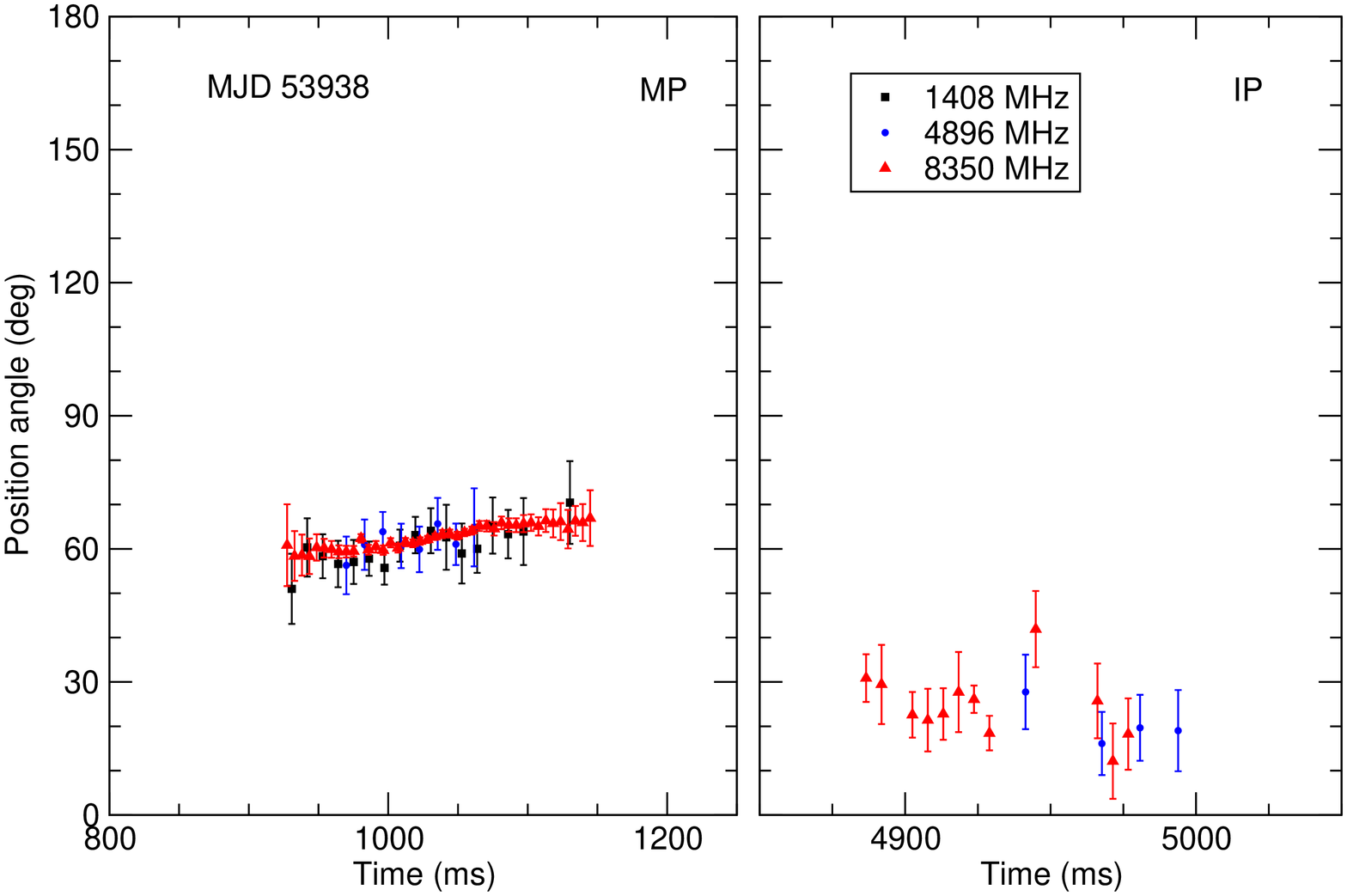,width=8cm,angle=-90} \\ 
\psfig{file=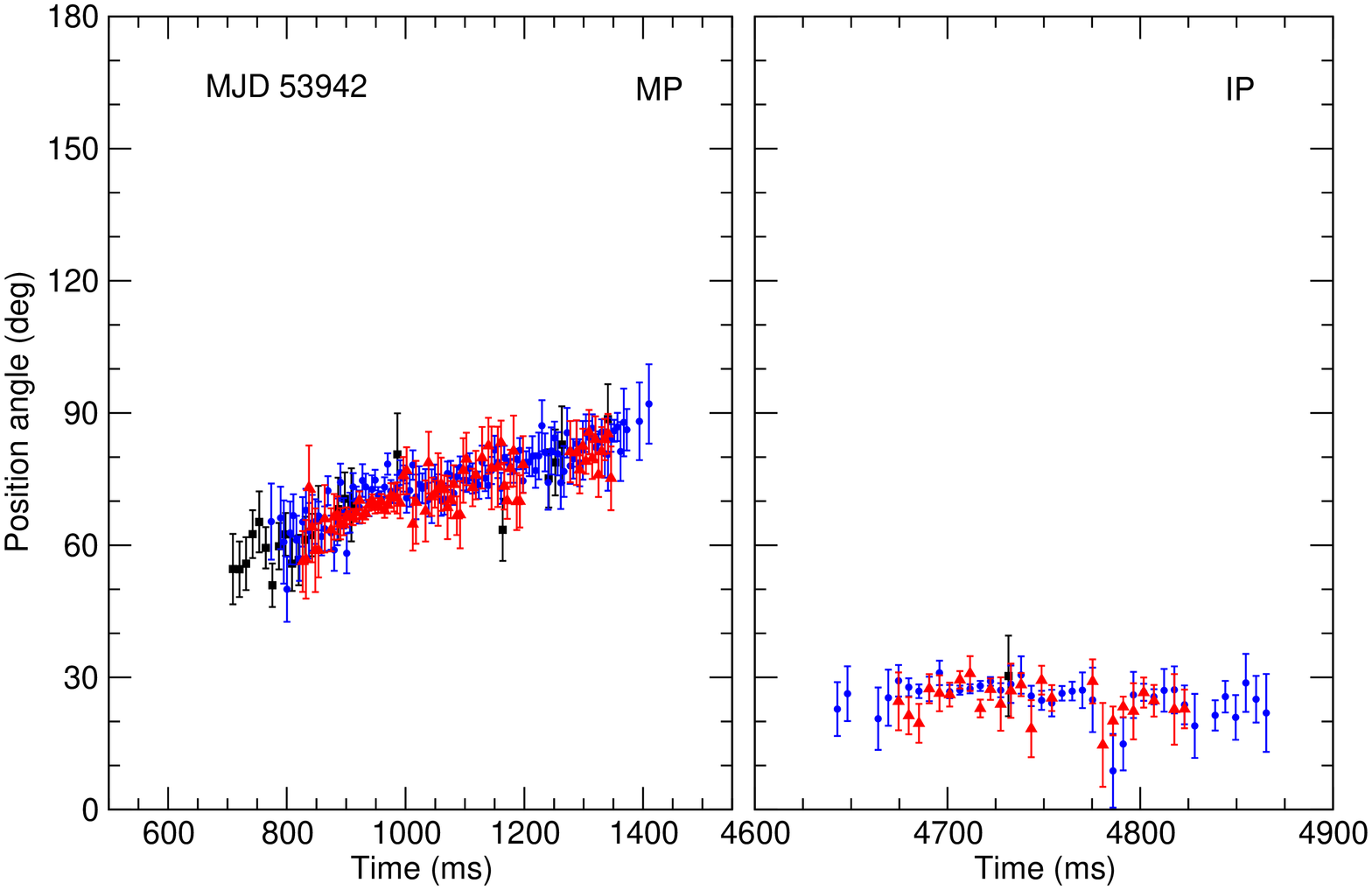,width=8cm,angle=-90} & 
\psfig{file=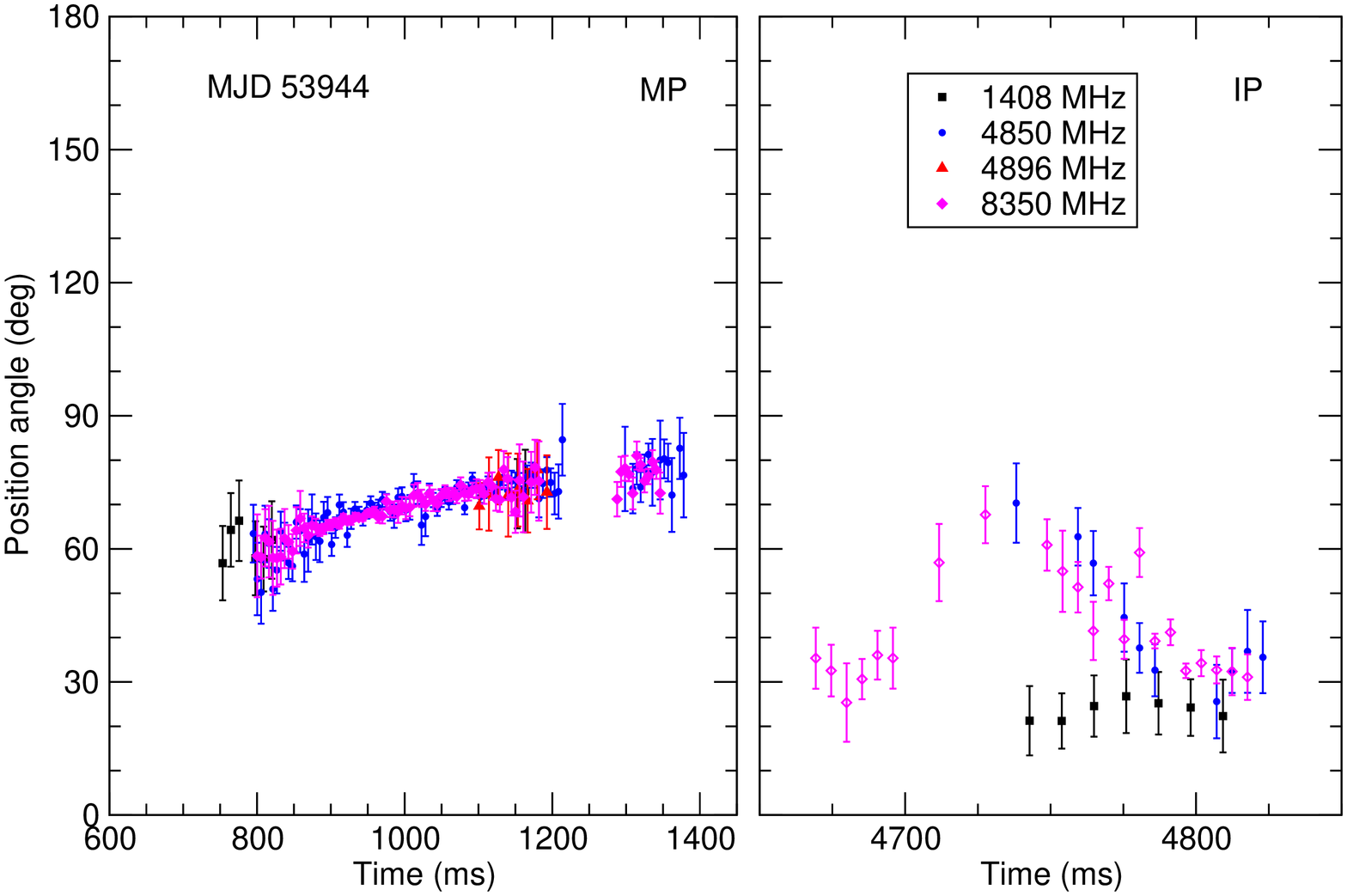,width=8cm,angle=-90} 
\end{tabular}
\end{center}

\caption{\label{fig:pa1}
Position angle swing of the linearly polarized components
observed at  different frequencies for several observing sessions,
namely 1, 2, 3, 5, 7 and 8.
}
\end{figure*}

Despite the striking changes in average pulse profiles presented
above, the PA swing in the MP is changing much less dramatically and
in a more organized fashion. The situation is quite different for the
IP. In order to verify that the PA changes seen in the MP are a
function of time, rather than observing frequency, we first compare
the PA swings seen simultaneously at different frequencies.  In
Figure~\ref{fig:pa1} we plot the PA at longitudes where the linear
polarized emission is strong enough to exceed a signal-to-noise ratio
threshold of 4. The agreement between the different frequencies is
extremely good, demonstrating the general broad-band character of the
PA swing, while simultaneously validating our calibration
procedures. Note that the agreement is found for both the MP and IP,
with the only exception being the IP data measured at 1.4 GHz during
session 8.

Having calibrated the absolute position angles for both 4.9 GHz and 8.4 GHz,
we use a measured offset of $\Delta PA = -10^\circ\pm2^\circ$ to infer a
rotation measure of RM$=+71\pm14$ rad m$^{-2}$.

\begin{figure*}

\begin{center}
\begin{tabular}{cc}
\psfig{file=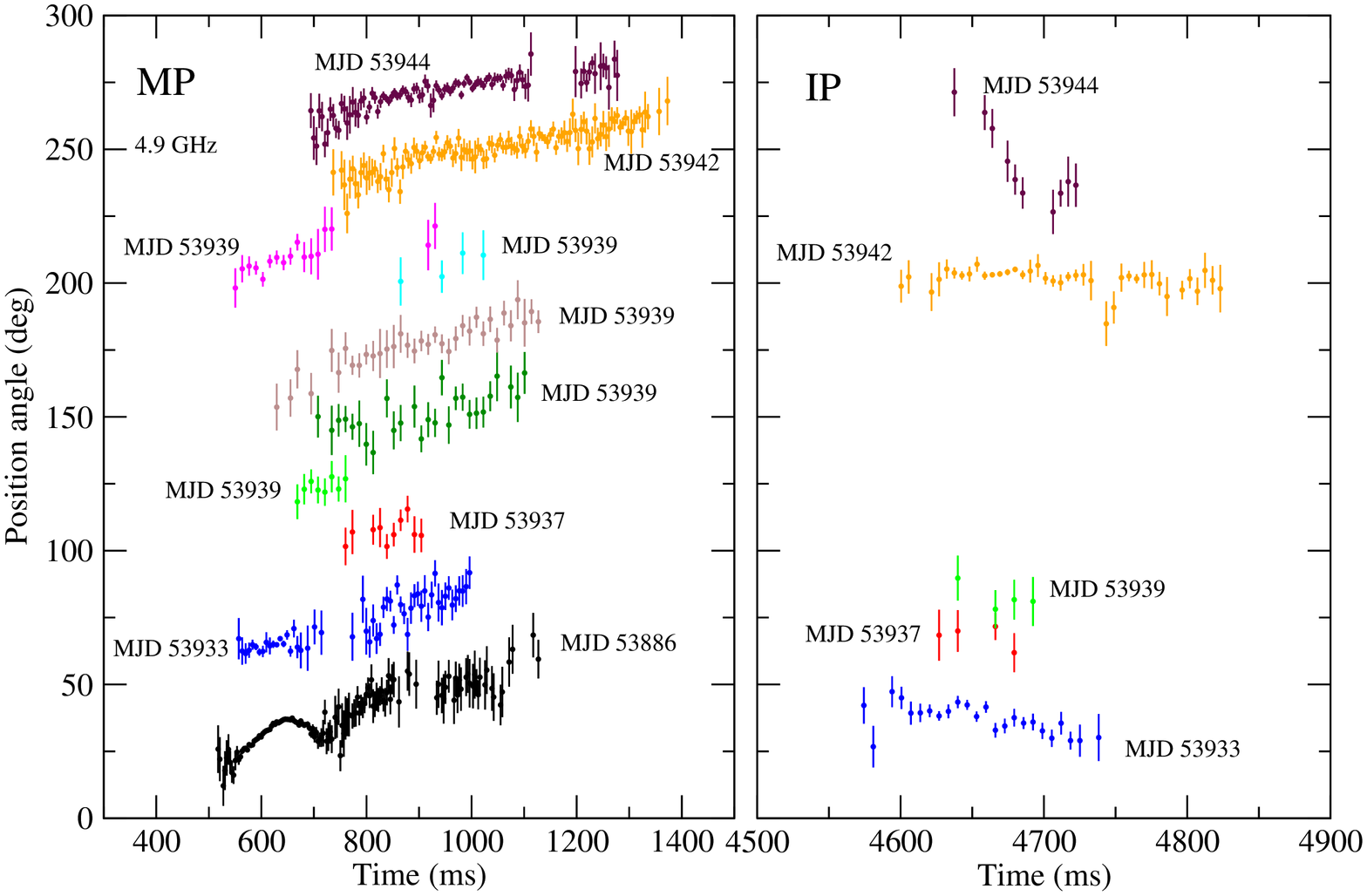,width=8cm,angle=-90} & 
\psfig{file=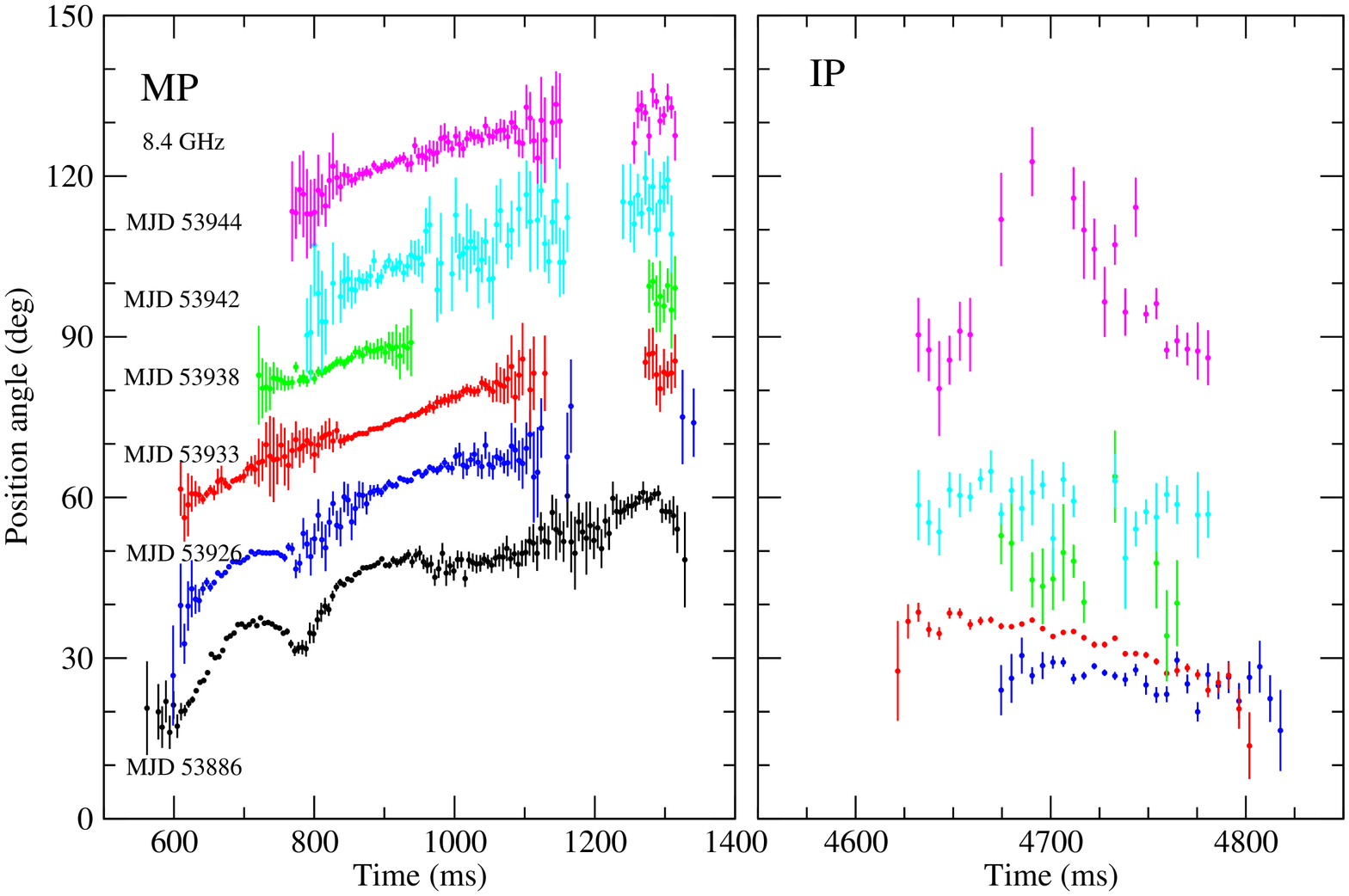,width=8cm,angle=-90} 
\end{tabular}
\end{center}

\caption{\label{fig:pa2}
Position angle swing as measured at 4850 MHz and 8350 MHz as
a function time for the MP and IP components.
}
\end{figure*}

As was evident from a comparison of the PA swings in
Figure~\ref{fig:align}, the PA changed between the observing sessions. In
Figure~\ref{fig:pa2}, we show the evolution of the PA as a function of
time in more detail for both the MP and IP as measured for 4.9 GHz
and 8.4 GHz. For reasons of clarity, we have offset the PA swings to
each other by a fixed amount which is the same each for both the MP
and IP.  A clear trend is visible where the previously discussed dip
becomes less prominent from session 1 to session 2 before it has
disappeared at session 5.  Interestingly, the longitude range where
the dip was present is later replaced by PA values which show a
significant scatter, typically much larger than at other pulse
longitudes. A slight change in the overall slope is also visible.  The
figure also demonstrates again how the leading MP component disappears
with time, while the most trailing component of the MP remains
essentially detectable at all times, at least at 8.4 GHz.
Surprisingly, observations at both frequencies with both Effelsberg and
Westerbork show an interesting 'wiggle' or oscillation in parts of the
PA swing which follows the longitudes of the 'dip' seen in session 1.

When studying the evolution of the PA swing measured in the IP,
it is obvious that it changes significantly with time. In addition
to obvious changes in the swing, it is remarkable that it also changes
in absolute PA value. This is notable by comparing the separation of the
PA values measured at different days and relating this to the fixed
offset introduced to the PA curves and visible in the MP plots. As
the PA swings in the MP components agree perfectly as shown above,
this variation in absolute PA is clearly intrinsic to the source.
A possible explanation for this effect is obtained when considering
the existence of polarization modes.

\subsection{Polarization modes}

Normal radio pulsars are known to exhibit linear polarisation modes which are
typically orthogonal to each other. The origin of these so-called orthogonal
modes (e.g.~Backer et al.~1976)\nocite{brc76} 
is believed to be related to mode-separating birefringence effects
in the pulsar magnetosphere (McKinnon 1997, Petrova 2001).\nocite{mck97,pet01} 
Simultaneous multi-frequency observations
found (Karastergiou et al.~2002)\nocite{kkj+02} that there is a high degree of
correlation between the polarization modes at two different frequencies.  They
also found that the modes occur more equally towards higher frequencies,
providing some explanation for the de-polarization of pulsar emission at high
frequencies, as overlapping orthogonal modes lead to a lower average degree of
polarization. 

\begin{figure*}

\begin{center}
\begin{tabular}{cc}
\psfig{file=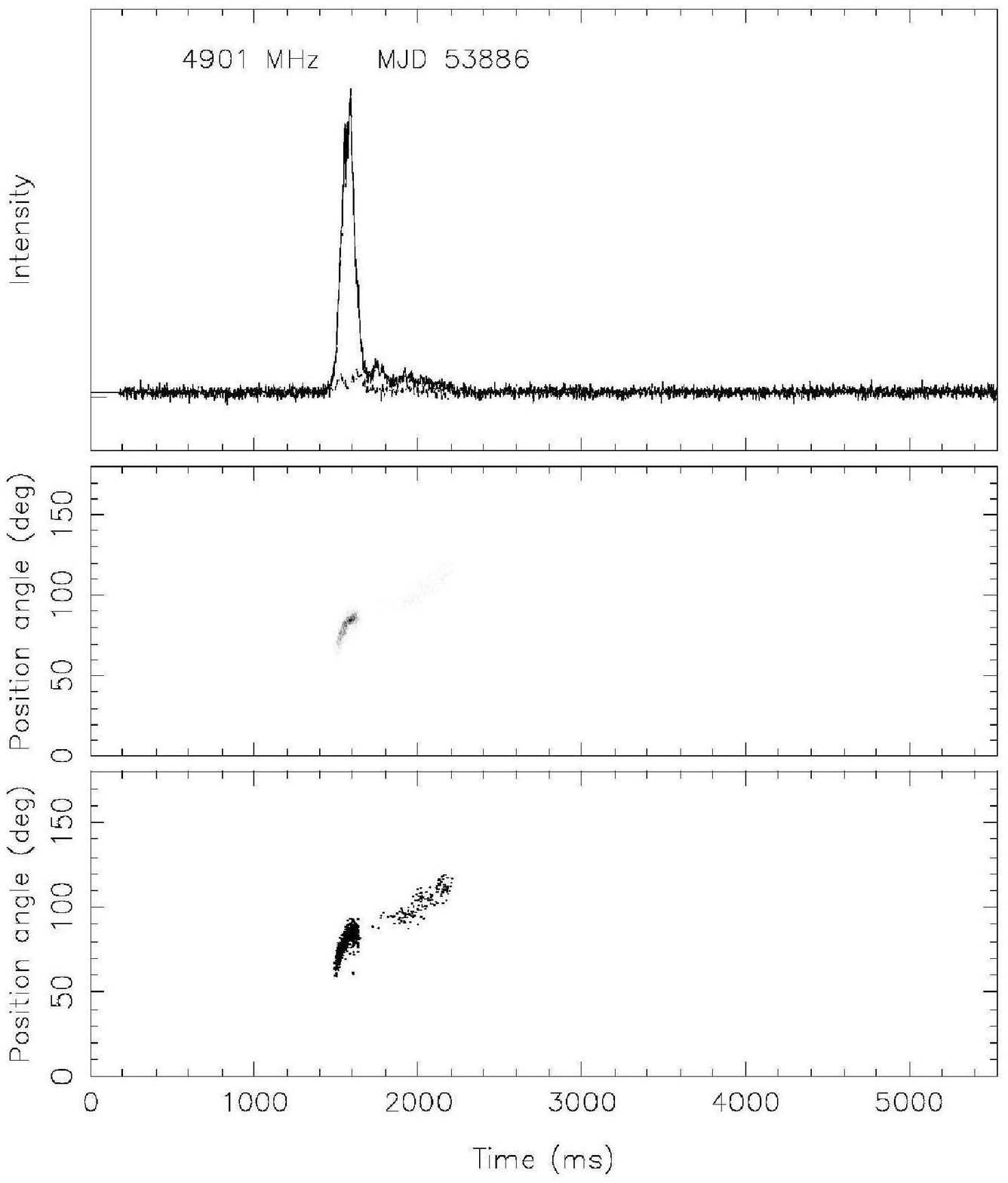,width=6cm} & 
\psfig{file=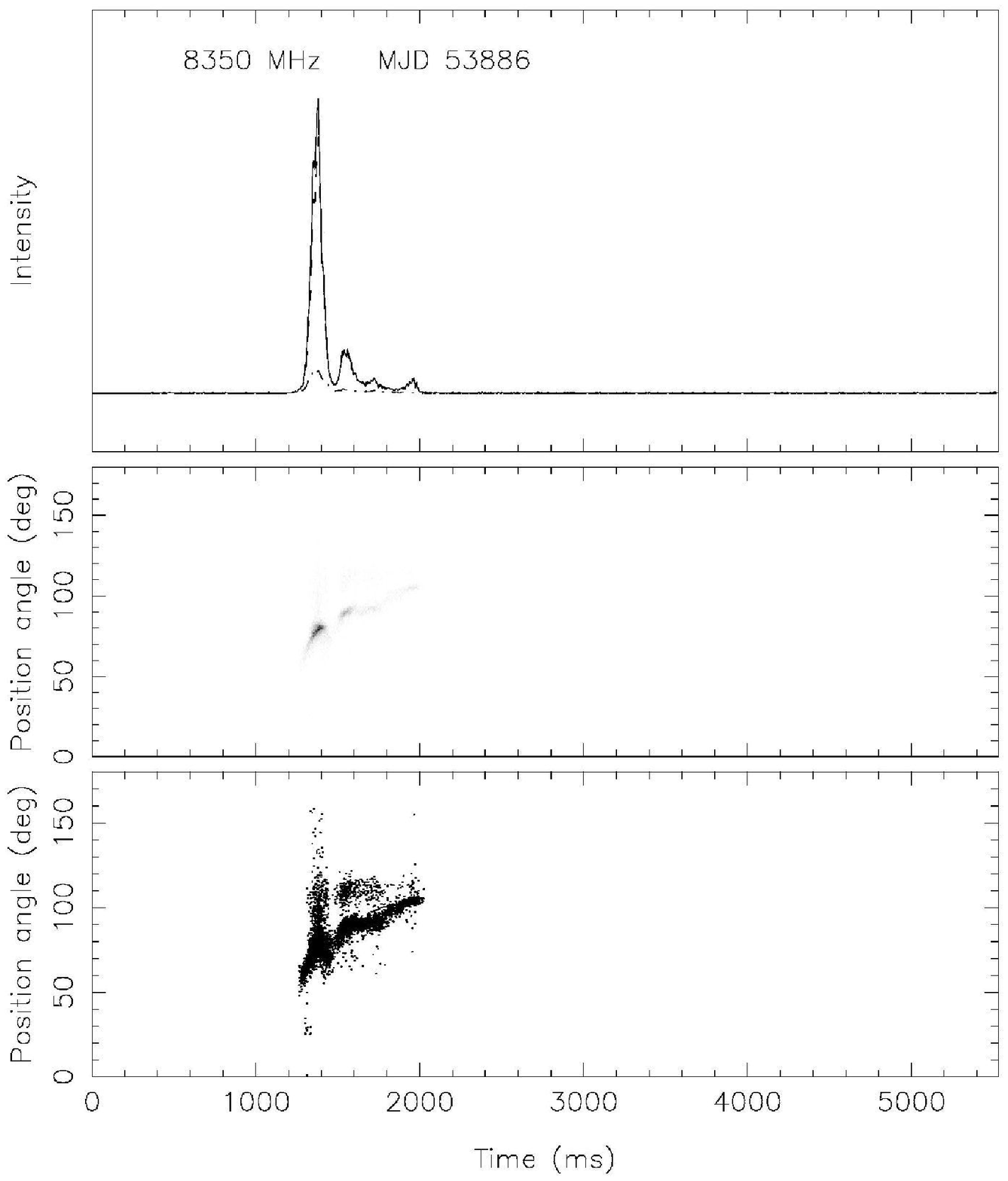,width=6cm} \\
\psfig{file=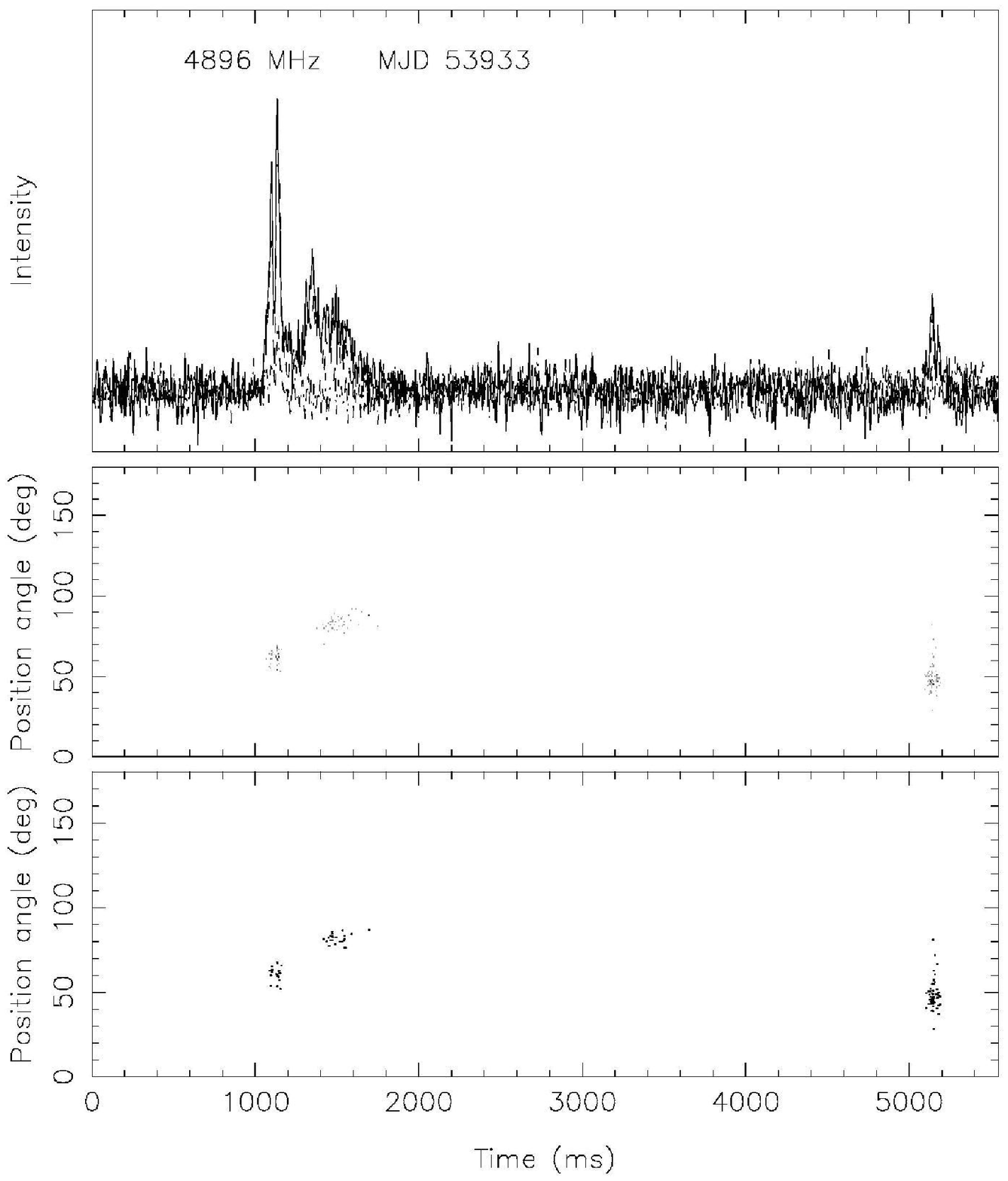,width=6cm} & 
\psfig{file=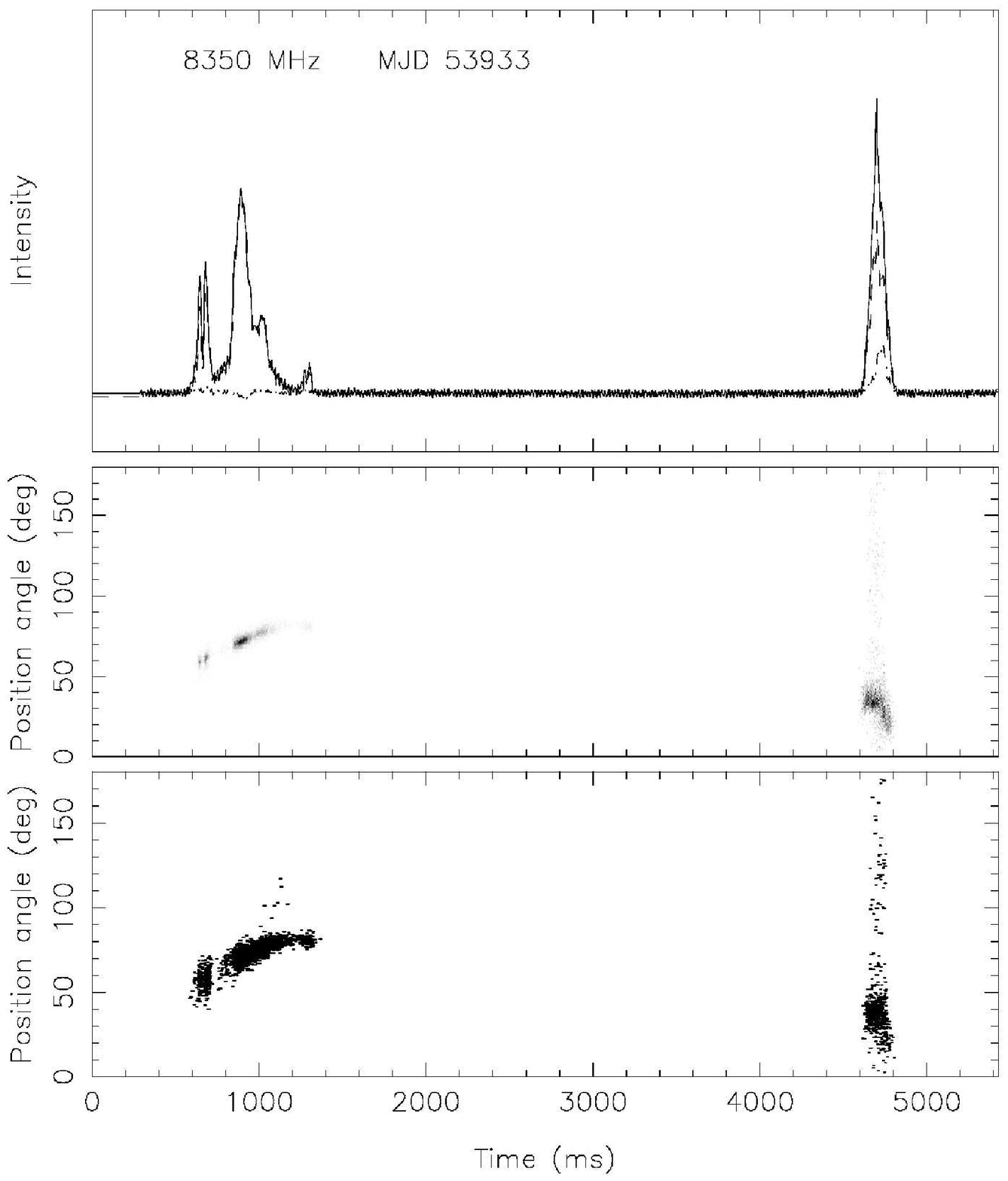,width=6cm} \\
\psfig{file=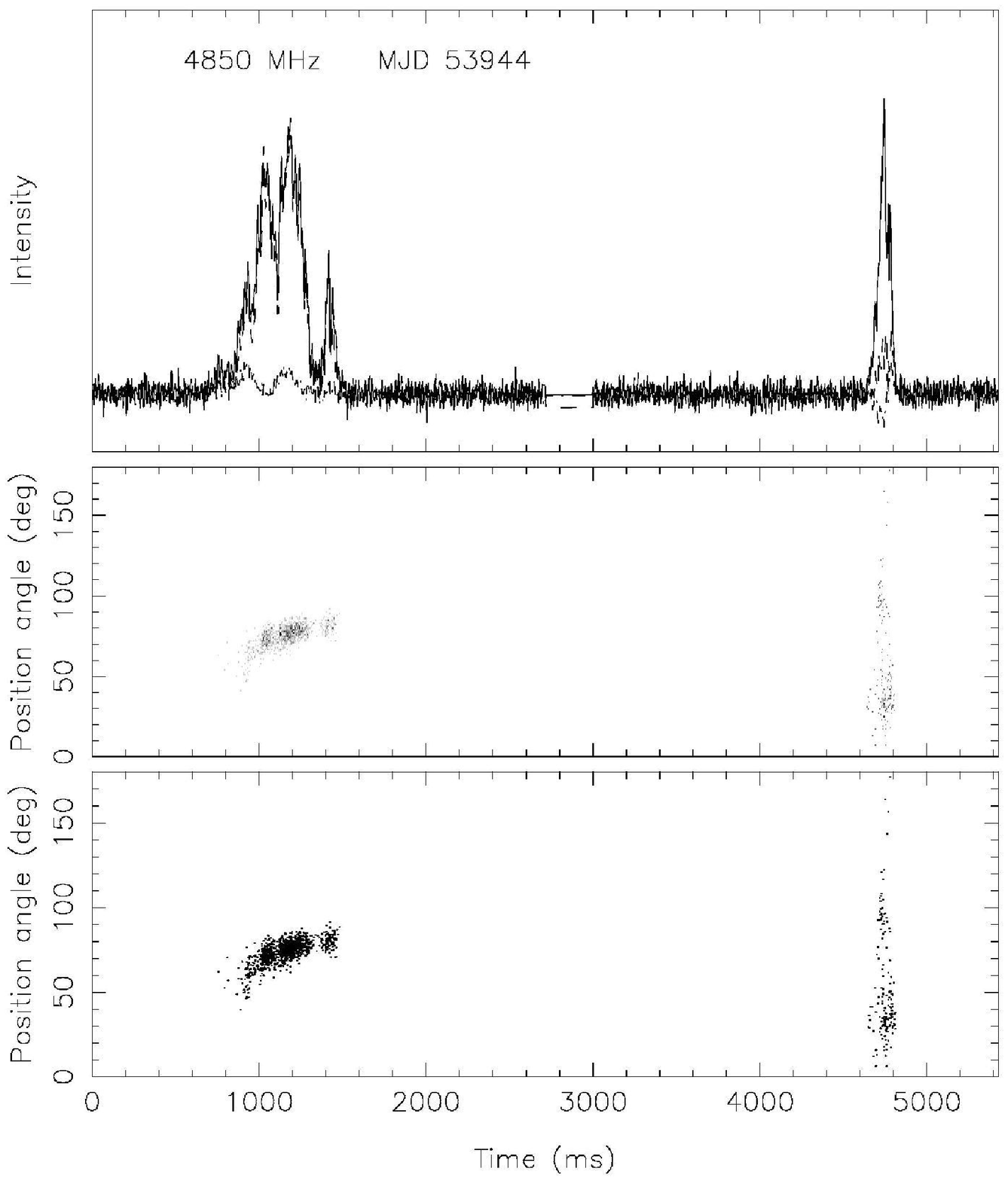,width=6cm} & 
\psfig{file=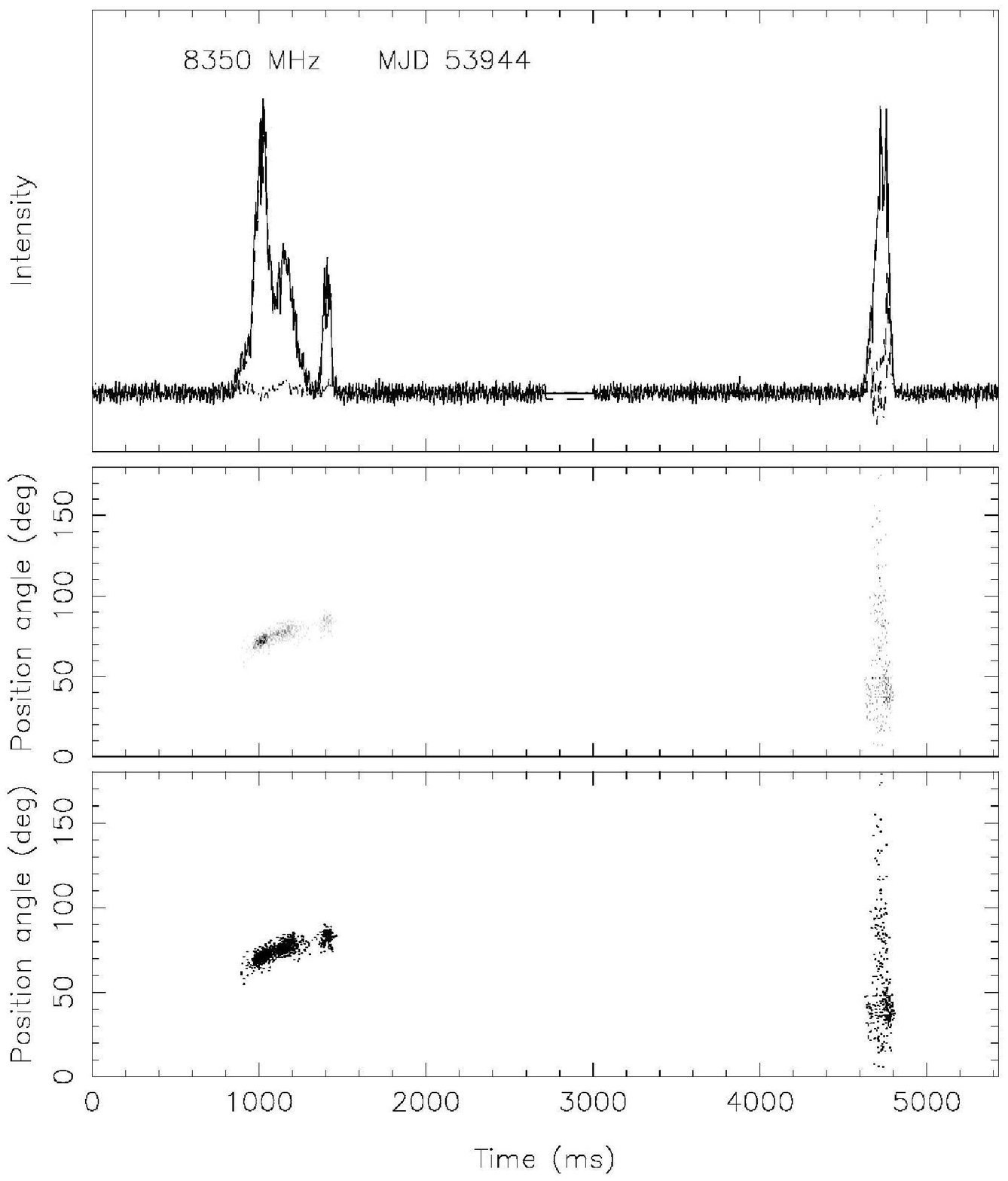,width=6cm} 
\end{tabular}
\end{center}

\caption{\label{fig:pahist}
  Distribution of occurrence of the position angles of the linearly polarized
  components measured at 4.9 GHz and 8.4 GHz for individual pulses during
  sessions 1, 3 and 8.  The top panel in each plot shows the average pulse
  profile with linear and circular intensity. The middle plot indicates how
  often a particular PA value was measured per phase bin, using a intensity
  scale where dark regions mark the highest occurrences. The MP and IP regions
  were scaled independently, so that the darkest shades of gray correspond to
  the most frequently measured PAs in each region respectively.  The bottom
  plot shows a scatter plot, indicating simply the measured PA value,
  irrespective of the frequency of occurrence.  }
\end{figure*}

The extremely high degree of polarization seen at all frequencies for
\axp{} would suggest that different orthogonal modes should not be
present as they often lead to a de-polarization of the average pulse
profiles (McKinnon 1997, Karastergiou et al.~2002). In order to verify
this expectation, we studied the distribution of PAs for each phase
bin.  To avoid spurious contributions, we only registered those
signals where both the total and linear intensity were 6 times
stronger than the corresponding off-pulse RMS. The resultant PA
histograms for three epochs are plotted in Figure~\ref{fig:pahist}
where we concentrate on the highest S/N data, i.e.~frequencies of
4.9 GHz and 8.4 GHz.

For each epoch and frequency, we show two PA plots. The bottom plot shows a
scatter plot, indicating simply the measured PA values, irrespective of the
frequency of occurrence. The middle plot shows instead how often a particular
PA value was measured per phase bin, using a gray-scale plot.  The darkest
regions mark the highest occurrences of PA.  The MP and IP regions were scaled
independently, so that the darkest shades of gray correspond to the most
frequently measured PAs in each region respectively.

In the histogram for session 1 at 8.4 GHz, one can see two well separated
regions of highly populated PA values indicating the existence of two
polarization modes. The two observed modes are non-orthogonal and are present
at nearly all pulse longitudes - apart from the 'dip' region where only one
mode seems to be present.  The reason why this mode separation does not lead
to a complete de-polarization of the average profile becomes clear when
studying the middle panel, which reflects the occurrence of the PA values.
While the existence of the two modes is highly significant, the second mode
does not occur very frequently. Interestingly, studying the PA values at 4.9
GHz, the second mode is not visible despite this data set covering a longer
time span, containing 1437 pulses, compared to 972 pulses obtained at the
higher frequency. In contrast, it is important to note that the average pulse
profiles at both frequencies are in perfect agreement. The fact that we also
observe circular polarization in the same pulse longitude range may therefore
indicate that we observe a conversion from linear into circular power at high
frequencies as the result of a propagation effect.

The occurrence of a second polarization mode seen in the MP is also changing
between observing session. In the histograms shown for session 3, the MP
polarization is essentially confined to one mode and the spread within this
mode is also smaller. However, in contrast, the polarization under the IP
shows an extremely large scatter in the PA swing. On the one hand, a very wide
range of PA values occurs as seen in the bottom panels. On the other hand,
even though PA values that differ greatly from the average PA are not very
common (see middle panels) the scatter around the average values is still
quite large. This explains why the shape and absolute angle of the IP seem to
change with time as the average IP properties then depend on the particular
single pulses added in the IP region. Although the signal-to-noise ratio at 4.9
GHz is worse than at 8.4 GHz, an identical behaviour is clearly seen at
both frequencies.

The third epoch shown Figure~\ref{fig:pahist} is one where the 'wiggle' in the
PA swing discussed earlier are very clearly seen. And indeed, both the scatter
plot as well as the PA density plot clearly indicate a spread and almost
bimodal distribution in the corresponding pulse phase range.  The regularity
of the wiggle variation is quite striking and may suggest the existence
of propagation effects in the magnetosphere that can change the ratio of
Stokes $Q$ and $U$ in a systematic way as a function of pulse longitude.

\begin{figure}

\begin{center}
\psfig{file=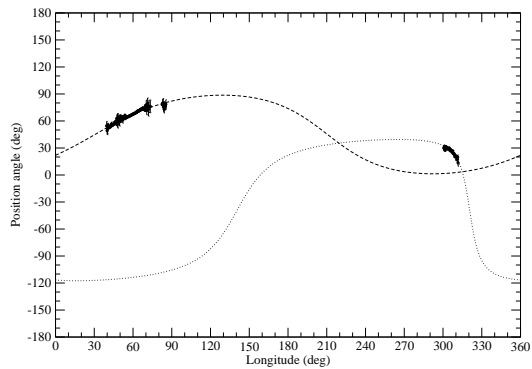,width=8cm,angle=-90} 
\end{center}

\caption{\label{fig:rvm} Rotating vector model fits for the 8.4-GHz
data measured in session 3. Two separate rotating vector models are
fitted to the MP and IP separately as no acceptable solution exists
for a simultaneous fit.  The fits shown produce an identical viewing
angle of $\alpha + \beta =83^\circ$.}
\end{figure}

The occurrence of significantly different PA values in the IP region is also
seen at all epochs and explains why we find it difficult to fit an acceptable
rotating vector model \cite{rc69a} to both the MP and IP polarization data.
Such a fit can in principle produce a value for the magnetic inclination angle
$\alpha$ and the impact angle of the line-of-sight to the magnetic pole
$\beta$ (e.g.~Lyne \& Smith 2006). A number of models, including wide cones
and dual pole models with and without orthogonal jumps between the IP and MP
were tried and no simultaneous fit to both sets of PAs were possible at any
frequency. Instead, one can find independent fits to the MP and IP where at
least the viewing angle, i.e.~ the angle between the rotation axis and the
line-of-sight, are consistent. An example for such a solution is shown in
Figure~\ref{fig:rvm} for the 8.4 GHz data of session 3.  Here we find for the
MP data a solution ($\alpha = 44^\circ\pm1^circ$, $\beta=39^\circ\pm4^\circ$)
and for the IP ($\alpha = 76.6^\circ\pm0.2^\circ$, $\beta=6^\circ\pm4^\circ$).
As indicated, both solutions have an identical value for $\alpha + \beta =
83^\circ$. We discuss the interpretation of this solution further below.

\label{sec:rvm} 

\subsection{Stability of polarization}

\begin{figure}
\centerline{
\psfig{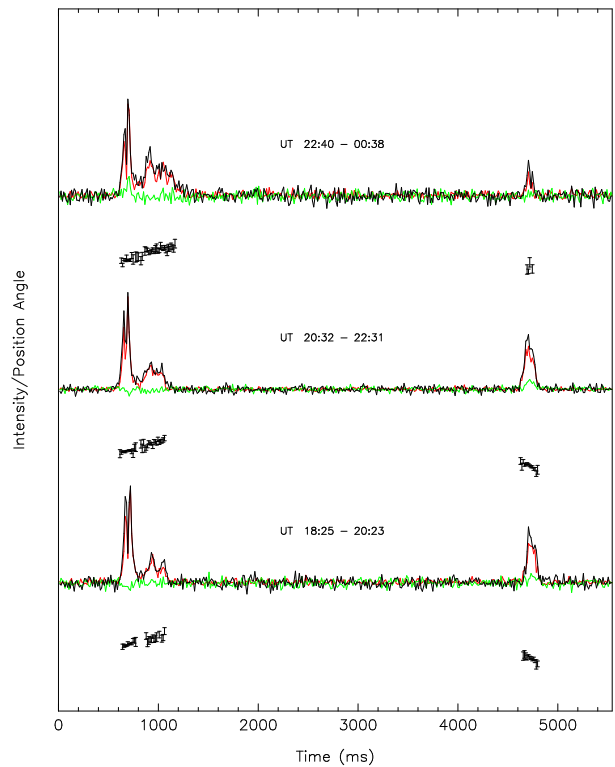} 
}

\caption{\label{fig:changingV} Example of short-time variation in the
circular polarization of the average pulse profile. The shown observations
were recorded at 4896 MHz in the night from MJD 53933 to 53934. Each
profile corresponds to about 120 min of integration time. The change in
the handedness of the circular polarisation in the MP  is clearly visible.
At the same time, the circular polarization in the IP remains essentially
unchnaged.
}
\end{figure}

As indicated earlier, apart from a long-term evolution of the pulsar profile
and its polarization properties, we also see short-term variations, including
variations in the circularly polarized emission component. In
Fig.~\ref{fig:changingV} we show an example of WSRT data taken during session
3 over a period of 6 hours. While the circular polarization of the IP region
seems to be unchanged within the uncertainties, the circular polarization in
the MP shows changes from negative to positive handedness. A similar behaviour
is also seen on long time-scales with the MP showing changes even though the
average degree of circular polarization remains low and mostly positive.  In
contrast, the IP tends to be more stable in circular polarization but in
particular in the later sessions stronger variations are observed (see
Fig.~\ref{fig:align}).

\section{Discussion}

It is clear that the radio emission of \axp shows several features
observed in ordinary radio pulsars, but also shows a large variety of
striking and peculiar differences. Among the most striking differences are
probably the flat spectrum of \axp, making it an extraordinarily
strong source at high radio frequencies, and the constantly changing,
extremely highly polarized pulse profile. 

In normal radio pulsars, the addition of several hundred pulses is
usually sufficient to reach a stable pulse profile (Helfand et
al.~1975\nocite{hmt75}, Rankin \& Rathnasree 1995\nocite{rr95}).  On
short time-scales of the order of minutes to hours, changes are
sometimes seen
as a phenomenon known as mode-changing (Backer 1970\nocite{bac70a})
in which the pulsar switches between a small number of
(typically two) stable pulse profiles. The dramatic changes between a
large variety of pulse shapes on time-scales ranging from minutes to
weeks as seen in \axp are clearly unusual. Perhaps, this source does
not exhibit a stable pulse shape as such, or the stabilization
time-scale is extremely large. The latter would indeed be consistent
with the extreme narrowness of the very spiky single pulses being
observed.

Indeed, rather than having broad sub-pulses as typically seen in
normal pulses, the spiky pulses occur at seemingly random positions
within the range of pulse phases for which emission is detected in the
average pulse profile. Such spiky emission is only rarely observed in
normal pulsars, but when it is observed it also leads to very long
profile stabilization time-scales. Weltevrede et
al.~(2006b)\nocite{wwsr06} observe a time-scale of over 25,000 pulses
for PSR B0656+14 at 327 MHz. Scaling this time-scale to the period of
\axp, this would imply a required observing time of about 38 hours to
reach a stable pulse profile. This is much longer than the time taken
to obtain any of the profiles presented so far in this paper or by
other authors, making this possibility a plausible explanation. On the
other hand, the range of pulse phases for which emission is detected
also changes systematically over timescales of weeks, with the first
component apparently slowly disappearing at later epochs. This may
indicate that the situation is more complex and that a stable pulse
profile may indeed be nonexistent in the first place.

The spiky appearance of the single pulses of \axp is more reminiscent
of the narrow 'giant micropulses' found first for the Vela pulsar
(Johnston et al.~2001)\nocite{jvkb01} and later for a few others
(Johnston \& Romani 2002)\nocite{jr02}. These giant micropulses tend
to occur at the edges of pulse profiles, in contrast to the wide
variety of pulse phases for which the spiky pulses of \axp are
seen. Moreover, pulsars with giant micropulses often have relatively
large magnetic fields near the light cylinder, similar to pulsars
exhibiting 'real' giant pulses (see e.g.~Lorimer \& Kramer 2005).  In
contrast, the magnetic field at the light cylinder of \axp, as
inferred from the observed period and spin-down, is only about 14
Gauss -- several orders of magnitudes lower than those estimated for
Vela or the Crab pulsar.
Even though the evidence for the connection
between the detection of giant pulses and a strong light-cylinder
magnetic field is somewhat circumstantial, it suggests that the emission
mechanism creating giant pulses may not be directly related to that
creating the radio emission of \axp. However, it is plausible that the
pulses from \axp originate from a much lower height in the
magnetosphere where similar conditions, in particular larger magnetic
field strengths, would be encountered.  Unfortunately, we know 
little about the emission properties of RRATS (e.g.~their polarization
properties) at the moment to allow for a similar, meaningful
comparison with these sources. It will be intriguing to see whether
the hypothesis that RRATS provide a link between radio pulsars and
magnetars (McLaughlin et al.~2006) is confirmed by finding similar
emission properties to be determined in future RRATS studies.

While the total power pulse profiles differ significantly between the
frequencies, the same applies to the linearly polarized components, as the
degree of polarization typically exceeds 90\% at most MP longitudes, following
the total power profile closely. In contrast, the degree of linear
polarization for the leading MP component is lower and significant amounts
for circular polarization are observed. For the rest of the MP the degree of
circular polarization remains low. Some changes are
observed but as far as it can be deduced from the generally much lower
signal-to-noise ratio they are much less dramatic than those of the linear or
total power profile.

Apart from a clear evolution over a timescale of weeks, one of the most stable
features in the pulse profile is the PA swing of the MP. There is also
excellent agreement between the PA seen at different frequencies when detected
at the same pulse longitudes.  This agreement appears to be even better than
the general good agreement found for pulsars by Karastergiou \& Johnston
(2006).\nocite{kj06} 

If the PA swing is interpreted geometrically as in the rotating-vector-model,
a change in the PA should only occur if either the geometry is changing
(e.g.~due to precessional effects) or if the underlying magnetic field
structure is undergoing changes. Stability of the magnetic field structure is
usually considered to be a safe assumption for radio pulsars where the
long-term stability of the average pulse profile and its position angle swing
is attributed to the dominance and strength of the dipolar magnetic field
(e.g.~Manchester \& Taylor 1977\nocite{mt77}). Even though the PA swing of
normal pulsars often shows deviations from an S-like shape (Lyne \& Manchester
1988\nocite{lm88}) and sometimes exhibits some variation across frequencies,
this is often attributed to the existence of orthogonal emission modes and
their different spectral properties (Karastergiou \& Johnston 2005).  The
long-term evolution of PA swing in the MP of \axp is therefore highly unusual.
The clear trend seen over timescales of days and weeks suggests that this is a
relatively slow process. Distinguishing between the possibilities as to
whether this is related to changes in the magnetic field structure, or to
precessional effects, or to the fact that the PA swing is not reflecting the
geometry of the system after all and that propagation effects in the
magnetosphere or non-dipolar field components are involved, is difficult.
However, there are a number of observed properties which indeed suggest the
existence of severe propagation effects in the magnetosphere: (a) we see
distinct polarization modes which change with radio frequency and time; (b) we
see a large variety of PA values in the IP region; (c) we have indications of
a conversion of linear into circular power for certain pulse longitude ranges
(d) we see interesting wiggles in the PA slopes which are hard to explain by a
any geometrical model and (e) even the changing flux density spectrum of the
individual MP and IP components could be due to propagation effects.  Indeed,
the magnetosphere is much larger than that of a typical radio pulsar, and the
range of inferred magnetic field strengths encountered from the surface
($B=2.6\times10^{14}$ Gauss) to the light-cylinder ($B=14$ Gauss) is certainly
enormous, covering 13 orders of magnitude and hence giving scope for a large
variety of effects.

Despite the likelihood of magnetospheric propagation effects, it is tempting
to associate the deviations from an S-like swing (and the inability to find a
single set of rotating-vector-model
parameters that connects both the MP and IP with a
satisfactory fit) with deviations from a dipolar field structure. We also
note that the separation of the IP from the MP is quite different from
the 180 deg expected in a two-pole model, and that the emission properties of
the IP are rather different from those of the MP, supporting the
interpretation of detecting signatures of non-dipolar field lines. Indeed,
within the magnetar model a dipolar field structure is not necessarily
expected. Similar arguments have been put forward to explain some of the
polarization and profile properties of millisecond pulsars (e.g.~Xilouris et
al.~1998)\nocite{xkj+98} where the evolutionary history of the sources may
lead to non-dipolar and sun-spot-like field structures (Ruderman
1991)\nocite{rud91}. However, depending on the emission height and hence the
physical separation from the field's multipole components, their actual impact
in the emission region may actually be low, although they may affect the
plasma flow from the surface and hence the observed radio properties. 

The relative shallowness of the PA swing under the MP with a slope of only 1
deg/deg is not uncommon for pulsars whereas the PA evolution with epoch
certainly is. It is interesting to compare the properties of \axp with those
of young pulsars, as the characteristic age of \axp is less than 10,000 years.
Johnston \& Weisberg (2006)\nocite{jw06} studied 14 young pulsars with
characteristic ages less than 75 kyr and found that generally pulse profiles
are simple and consist of either one or two prominent components whereas the
linearly polarized fraction is nearly always in excess of 70 per cent. The
latter characteristic is certainly true for \axp but the profile is clearly
anything but simple, nor does the trailing component dominate, as Weisberg \&
Johnston (2006) also find, on average, for young pulsars.  

The flat PA curve could be both explained either by an aligned
configuration in which the spin-axis is aligned with the magnetic field
axis, or by a very wide cone which is cut far away from the
magnetic pole.  The solution for the RVM fits to the MP presented in
Section~\ref{sec:rvm} corresponds to an extremely wide cone grazed at
the outside for the MP ($\alpha=44^\circ$, $\beta=39^\circ$), with
a beam radius inferred from the MP pulse width of about $\rho\sim
44^\circ$. (Note that both the MP and IP are not centred on the
steepest gradient of the fitted RVM curves.) In contrast, the inferred
beam radius for the IP is much smaller, $\rho\sim8^\circ$. However,
there are some hints of a correlation in strength between the IP and
trailing MP components, so that one could consider both emission
features as part of another wide cone.  In comparison, the radio beam
of normal radio pulsars appears to scale with $\propto 1/\sqrt{P}$,
which would imply a beam radius for \axp of less than 3 deg when measured at a
50\% intensity level (Lorimer \& Kramer 2005). Alternatively, with an
aligned configuration, the observer's line-of-sight may hardly ever
leave the emission cone, also creating a wide pulse width.  
In any case, if we interprete the RVM fits geometrically, then we have to
conclude that we observe two emissions cones, that are centred on different
independent magnetic poles separated by 109 degrees in neutron-star longitude.
We can interprete that either as an offset dipole or evidence of a non-dipolar
field configuration. The viewing geometry and the rather different emission
properties of the various pulse components are certainly consistent with this
view.

It is intriguing that \axp shows variations in its spin-down (Camilo et
al.~2006b), which could be related to changes in the torque and hence the
magnetic field structure near the light-cylinder. If these torque changes are
related to the magnetic field structure changing near the light-cylinder, it
may also show changes in the emission region. In the magnetar model (Duncan \&
Thompson 1992)\nocite{dt92a} one expects that changes in the magnetic field
may trigger energetic outbursts. This seems to be inconsistent with the
observation that no X-ray variation was seen in recent monitoring observations
(Camilo et al.~2006b). However, Beloborodov \& Thompson (2006)\nocite{bt06}
propose that a plasma corona forms around the neutron star, as a result of
occasional starquakes that twist the external magnetic field of the star and
induce electric currents in the closed magnetosphere. If such effect is
present, one could speculate as to whether a combination of twisted fieldlines and
changing plasma corona is responsible for the observed emission properties.
Continuing studies of simultaneous multi-frequency data offer a chance to
indeed separate changing geometrical/field-configuration effects from
propagation effects. Further such studies are in progress and will be presented
elsewhere.

\section{Summary}

We have observed the magnetar \axp in full polarisation using three different
telescopes simultaneously at three different frequencies. We find that some
properties of \axp's are similar to those of normal radio pulsars while many
more features are observed that are strikingly different. We find strong
evidence for propagation effects in the magnetosphere while the observed
emission properties are consistent with a multi-pole configuration of the
magnetic field.  Continued observations of this radio emitting magnetar,
together with the future studies of RRATS sources, will allow us to study a
variety of neutron stars in the radio regime and to contrast their emission
properties with those of normal pulsars.

\section*{Acknowledgements}

We thanks Gemma Janssen for help with the data acquisition and Patrick
Weltevrede for software and useful discussions. We thank H. Spruit and G.
Smith for useful discussions.


\end{document}